\documentclass[]{raa}            

\usepackage{graphicx,times}             
\usepackage{natbib}
\usepackage{mathtools}
\usepackage{amsmath}
\usepackage{color}
\usepackage{fnpct}
\usepackage[multiple]{footmisc}
\usepackage{multirow}
\usepackage{longtable}
\usepackage[hyphens]{url}
\newcommand{\zphot}{$z_\mathrm{phot}$}
\newcommand{\zspec}{$z_\mathrm{spec}$}
\newcommand{\dz}{{$\Delta z_\mathrm{norm}$}}
\newcommand{\zbias}{$\overline{\Delta z_\mathrm{norm}}$}
\newcommand{\zsigma}{$\sigma_{\Delta z_\mathrm{norm}}$}
\newcommand{\pout}{$P_\mathrm{3\sigma}$}
\newcommand{\pcon}{$P_\mathrm{0.15}$}

\usepackage{enumitem}
\setenumerate[1]{itemsep=0pt,partopsep=0pt,parsep=\parskip,topsep=5pt}
\setitemize[1]{itemsep=0pt,partopsep=0pt,parsep=\parskip,topsep=5pt}
\setdescription[1]{itemsep=0pt,partopsep=0pt,parsep=\parskip,topsep=5pt}

\begin{document}

  \title{Photometric redshifts and Galaxy Clusters for DES DR2, DESI DR9, and HSC-SSP PDR3 Data}

   \volnopage{Vol.0 (20xx) No.0, 000--000}      

   \author{ 
     Hu Zou
      \inst{1,2}
   \and Jipeng Sui
      \inst{1,2}
   \and Suijian Xue
    \inst{1}
   \and Xu Zhou
   \inst{1}
   \and Jun Ma
   \inst{1,2}
   \and Zhimin Zhou
   \inst{1}
   \and Jundan Nie
   \inst{1}
   \and Tianmeng Zhang
   \inst{1}
   \and  Lu Feng
   \inst{1}
   \and Zhixia Shen
   \inst{1}
   \and Jiali Wang
     \inst{1}
   }


   \institute{Key Laboratory of Optical Astronomy, National Astronomical Observatories, Chinese Academy of Sciences,
             Beijing 100012, China; {\it zouhu@nao.cas.cn}\\
        \and
             School of Astronomy and Space Science, University of Chinese Academy of Sciences, Beijing 101408, China\\
\vs\no
   {\small Received 20xx month day; accepted 20xx month day} }

\abstract{Photometric redshift (photo-z) is a fundamental parameter for multi-wavelength photometric surveys, while galaxy clusters are important cosmological probers and ideal objects for exploring the dense environmental impact on galaxy evolution. We extend our previous work on estimating photo-z and detecting galaxy clusters to the latest data releases of the Dark Energy Spectroscopic Instrument (DESI) imaging surveys, Dark Energy Survey (DES), and Hyper Suprime-Cam Subaru Strategic Program (HSC-SSP) imaging surveys and make corresponding catalogs publicly available for more extensive scientific applications. The photo-z catalogs include accurate measurements of photo-z and stellar mass for about 320, 293, and 134 million galaxies with $r<23$, $i<24$, and $i<25$ in DESI DR9, DES DR2, and HSC-SSP PDR3 data, respectively. The photo-z accuracy is about 0.017, 0.024, and 0.029 and the general redshift coverage is $z<1$, $z<1.2$, and $z<1.6$, respectively for those three surveys. The uncertainties of the logarithmic stellar mass that is inferred from stellar population synthesis fitting is about 0.2 dex. With the above photo-z catalogs, galaxy clusters are detected using a fast cluster-finding algorithm. A total of 532,810, 86,963, and 36,566 galaxy clusters with the number of members larger than 10 are discovered for DESI, DES, and HSC-SSP, respectively. Their photo-z accuracy is at the level of 0.01. The total mass of our clusters are also estimated by using the calibration relations between the optical richness and the mass measurement from X-ray and radio observations. The photo-z and cluster catalogs are available at ScienceDB (\url{https://www.doi.org/10.11922/sciencedb.o00069.00003})  and PaperData Repository (\url{https://doi.org/10.12149/101089}).  
}

   \authorrunning{Hu Zou et al. }            
   \titlerunning{Photo-z and galaxy clusters }  

   \maketitle

%
%
\section{Introduction}           
\label{sect:intro}
Our understanding of the formation and evolution of the universe owes a great deal to modern large-scale imaging and spectroscopic surveys. One of the important parameters to be measured for astronomical objects is redshift (or equivalently distance), which is crucial to explore the galaxy evolution and cosmology. Although spectroscopic observations can provide accurate redshift measurements, they are time-consuming and flux-limited. The techniques of estimating photo-z become more and more important and even indispensable to the successes for some wide and deep imaging surveys, such as DES \citep{DES2005}, HSC-SSP \citep{Aihara2018}, LSST \citep{Lss2009}, and Euclid mission \citep{Laureijs2010}. One of the photo-z applications is to detect galaxy clusters, which are also important scientific objects in the above surveys. Galaxy clusters have been formed on the cosmic web. They trace the large-scale structure and are ideal laboratories to study the environmental effect on galaxy formation and evolution. As the largest gravitationally bound systems in the universe, galaxy clusters have been effectively detected in large-scale imaging surveys \citep{Hao2010,Rykoff2014,Rykoff2016,Zou2021}. 

In January 2021, DES made the second public data release \citep[hereafter DES DR2;][]{Abbott2021}. This release covers a sky area of $\sim 5,000$ deg$^2$ in the south Galactic cap. The 10$\sigma$ $i$-band magnitude limit is about 23.8 mag. In August 2021, HSC-SSP announced the third public release \citep[hereafter HSC-SSP PDR3;][]{Aihara2021}. The PDR3 release covers about 670 deg$^2$ in the wide layer at the $5\sigma$ depth of $i \sim 26$ mag and more than 30 deg$^2$ in the deep/ultra deep layer at the $5\sigma$ depth of $i \sim 27$ mag. In January 2021, the imaging team of the DESI project published the ninth data release (hereafter DESI DR9). It covers a sky area of $\sim$ 20,000 deg$^2$ in both south and north Galactic caps. The $5\sigma$ magnitude limit is about $r\sim23.9$ mag.

DES has not published the catalogs of photo-z and galaxy clusters yet. Although the HSC-SSP data releases previous to PDR3 have photo-z products, the PDR3 photo-z catalogs are also not delivered. The PDR3 significantly increases the sky area with all five filters to the required depths relative to the previous releases. We have successfully applied a local linear regression algorithm to accurately estimate the photo-zs for galaxies from South Galactic u-band Sky Survey (SCUSS) and DESI legacy imaging surveys \citep{Gao2018, Zou2019}. The resulting photo-z accuracy is at the level of 0.02. With these photo-z catalogs, we also developed a new cluster-finding method to identify galaxy clusters. A total of about 20,000 in SCUSS and 540,000 clusters in DESI have been founded \citep{Gao2020, Zou2021}. 

In this paper, we will generate new photo-z catalogs specifically for DES DR2 and HSC-SSP PDR3 data and update our photometric redshifts for the DESI imaging surveys to the latest DR9 data. Meanwhile, based on these photo-z measurements, we will derive reliable stellar mass for galaxies and detect a large number of galaxy clusters. Combining all these data, we can substantially extend the mass and redshift coverages of both galaxies and galaxy clusters. The catalogs can be made publicly available immediately, which will be superbly useful for further sciences. The structure of this paper is organized as follows. Section \ref{sec:data} describes the photometric and spectroscopic data. Section \ref{photoz} presents the photo-z and stellar mass measurements. Section \ref{sec:cluster} shows the detection of galaxy clusters. Section \ref{sec:summary} gives a summary. Throughout this paper, we assume a $\Lambda$CDM cosmology with $\Omega_m=0.3$, $\Omega_\Lambda=0.7$, and $H_0=70$ km s$^{-1}$ Mpc$^{-1}$.

\section{Data} \label{sec:data}
\subsection{Photometric data and galaxy sample}
\subsubsection{DES DR2}
DES is an imaging survey of about 5000 deg$^2$ in the south sky \citep{DES2005}, using the wide-field Dark Energy Camera \citep{Flaugher2015} installed on the 4m Blanco telescope. The main goal of DES is to study dark energy via constructing the three-dimensional distribution of galaxies using photometric redshifts. The adopted photometric system includes five optical broad filters (i.e., $grizY$). The first data release, DES DR1, was published in 2018 \citep{Abbott2018}. It includes the observations taken in the first three years. The DES DR2 was publicly available in 2021\footnote{\url{https://des.ncsa.illinois.edu/releases/dr2}}. It includes the data products assembled over all six years of DES science observations \citep{Abbott2021}. The sky coverage with all five-band photometry is about 4900 deg$^2$.  The magnitude limits at S/N $=$ 5 for point sources are $g=25.4$, $r=25.1$, $i=24.5$, $z=23.8$ and $Y=22.4$.

Galaxies in DES DR2 are selected using the following criteria:
\begin{itemize}
\item[-] mag\_auto\_i\_dered $<$ 24 (magnitude cut for $i$ band)
\item[-] imaflags\_iso\_i $=$ 0 (good photometric flag in $i$ band)
\item[-] flags\_i $<$ 4 (good photometric flag in $i$ band)
\item[-] extended\_class\_coadd $>=$ 2 (galaxy type)
\end{itemize}
Note that all the photometric magnitudes used in the following of this paper are corrected for the Galactic extinction. Finally, we select 292,636,425 galaxies with $i<24$.  

\subsubsection{HSC-SSP PDR3}
HSC-SSP uses a wide-field imaging camera deployed on the 8.2m Subaru telescope to carry out a wide and deep imaging survey \citep{Aihara2018}. The survey includes three layers. The Wide layer is planned to cover about 1400 deg$^2$ in five broad bands of $grizy$. The Deep and UtraDeep layers will cover more than 30 deg$^2$ in the five broad-band filters and four narrow-band filters. The $5\sigma$ magnitude limits are one and two magnitudes deeper than the Wide layer, respectively. The latest data release of HSC-SSP is PDR3 \citep{Aihara2021}, which was publicly accessible in August 2021\footnote{\url{https://hsc-release.mtk.nao.ac.jp/doc/}}. This release increases the sky coverage with full five-band photometry by two times more than PDR2. We only consider the Wide layer in this paper. The $5\sigma$ depths for point sources in the Wide layer are $g=26.5$, $r=26.5$, $i=26.2$, $z=25.2$, and $y=24.4$. 

In HSC-SSP PDR3, we use the forced measurements in which common object centroids and shape parameters are used for photometry in all filters. Galaxies in PDR3 ``forced" catalogs are selected using the following criteria:
\begin{itemize}
\item[-] isprimary $=$ True (no duplicates)
\item[-] i\_cmodel\_mag $<$ 25 (magnitude cut in $i$ band)
\item[-] [ri]\_extendedness\_value $=$ 1 (galaxy type in both $r$ and $i$ bands) 
\item[-] [ri]\_cmodel\_flag = False  (good photometric flag in both $r$ and $i$ bands)
\item[-] [ri]\_extendedness\_flag = False (good classification in both $r$ and $i$ bands)
\item[-] [ri]\_pixelflags\_saturatedcenter (no saturated objects)
\end{itemize}
In this way, we select 133,554,787 galaxies with $i < 25$.

\subsubsection{DESI DR9}
The legacy imaging surveys of DESI consists of three independent optical surveys conducted by three teams using three different telescopes \citep{Dey2019}: the Beijing-Arizona Sky Survey \citep[BASS;][]{Zou2017}, Mayall $z$-band Legacy Survey (MzLS), and DECam Legacy Survey (DECaLS). The BASS uses the 2.3 m Bok telescope on Kitt Peak, Arizona to survey a sky area of 5000 deg$^2$ with $g$ and $r$ bands in the north Galactic cap. The MzLS covers the same area using the 4m Mayall telescope on Kitt Peak with $z$ band. The DECaLS uses the 4m Blanco telescope to take $grz$-band imaging over 9000 deg$^2$ along the Equator in both north and south Galactic caps. In addition, the DESI imaging team makes new coadds of WISE $W1$ and $W2$ observations and performs forced photometry on these near-infrared images. These imaging surveys provide optical and near-infrared photometric data that are mainly used for the target selections of the DESI spectroscopic survey.  The latest data release is DR9, which was published in January 2021\footnote{\url{https://www.legacysurvey.org/dr9/}}.  The optical-band depths at $5\sigma$ are about $g=24.7$, $r=23.9$, and $z=23.0$ mag. The WISE data contain all 6-year imaging and the $5\sigma$ depths are $W1=20.7$ and $W2=20.0$ in AB mag, which are 1 mag deeper than AllWISE\footnote{\url{https://wise2.ipac.caltech.edu/docs/release/allwise/}}. The sky area with all $grzW1W2$ photometry in DR9 is about 19,000 deg$^2$.

We select the galaxies are in DESI DR9 using the following criteria:
\begin{itemize}
\item[-] mag\_r $<$ 23 (model magnitude cut in $r$ band)
\item[-] type $!=$ PSF (galaxy type)
\item[-] fracmasked\_[g,r,z] $<$ 0.5 (clean photometric cuts)
\item[-] fracflux\_[g,r,z] $<$ 0.5 (clean photometric cuts)
\item[-] fracin\_[g,r,z] $>$ 0.3  (clean photometric cuts)
\end{itemize}
A total of 320,060,206 galaxies with $r < 23$ are retained. The magnitude cuts for the above three surveys are roughly selected according to S/N of about 10.

\begin{figure*}[thb!]
\centering
\includegraphics[width=1\linewidth]{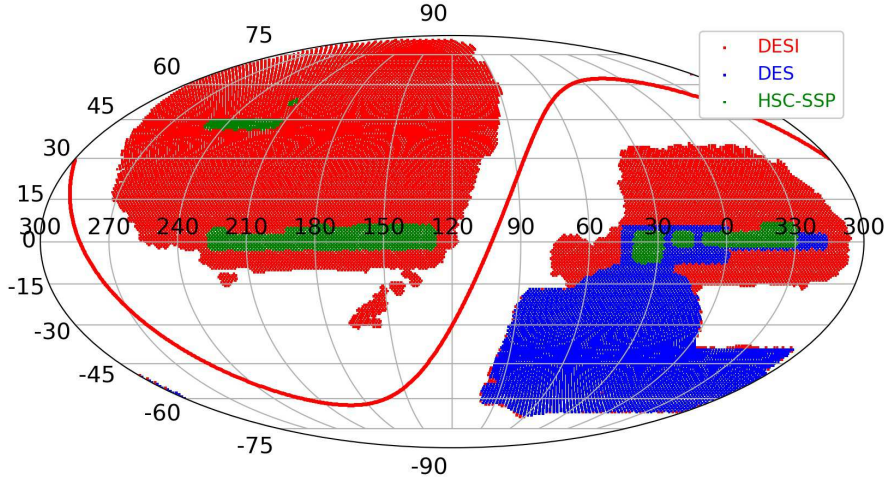}
\caption{Sky coverages of DESI (red), DES (blue), and HSC-SSP (green). \label{fig-coverage}}
\end{figure*}

Unless otherwise specified, hereafter we refer DESI, DES, and  HSC-SSP for short to DESI DR9, DES DR2, and HSC-SSP PDR3, respectively. Figure \ref{fig-coverage} presents the sky coverages of all above imaging surveys. The sky areas with full five-band photometry are 19,876, 5,194, and 1,128 deg$^2$ for DESI, DES, and HSC-SSP, respectively. Table \ref{tab-survey} summarizes the survey characteristics. 

\begin{table}
    \centering
    \caption{Sky coverage and imaging depths for different surveys\label{tab-survey}} 
    \begin{tabular}{c|c|c}
    \hline 
     \hline
        Survey & Sky area &  Depth$^a$   \\
            & deg$^2$  & mag   \\
      \hline
        DESI & 19,876  &  $g=24.7$, $r=23.9$, $z=23.0$, $W1=20.7$, $W2=20.0$ \\
        DES & 5,194 &  $g=25.4$, $r=25.1$, $i=24.5$, $z=23.8$, $Y=22.4$ \\
        HSC-SSP & 1,128 & $g=26.5$, $r=26.5$, $i=26.2$, $z=25.2$, $y=24.4$ \\
         \hline
\end{tabular}

{Notes: $^a$The depth here is referred to the $5\sigma$ limiting magnitude in AB mag.}
\end{table}

\subsection{Spectroscopic data}
\begin{figure*}[thb!]
\center
\includegraphics[width=\linewidth]{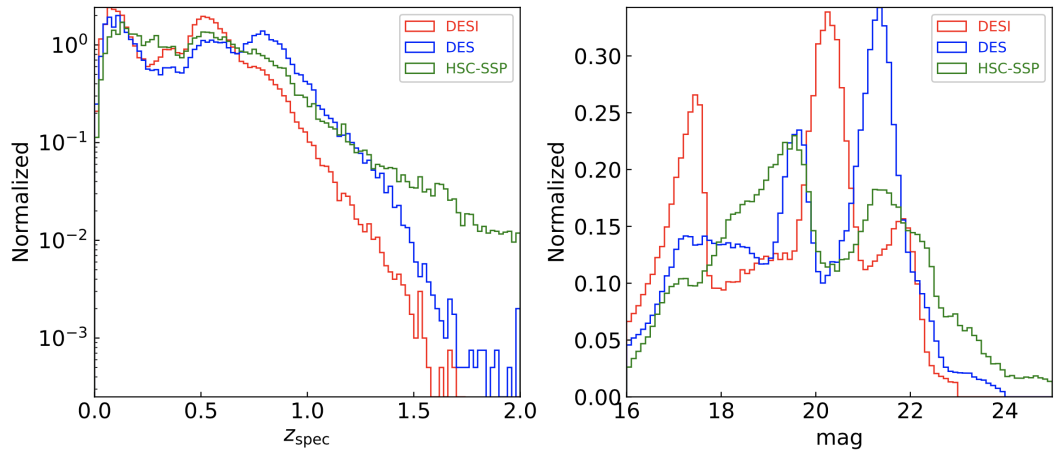}
\caption{Left: normalized distribution of the spectroscopic redshift for the training samples. The y axis is plotted in logarithmic to highlight the high-redshift end. Right: normalized distribution of the magnitude ($r$ band for DESI and $i$ band for others). \label{fig-zprop}}
\end{figure*}

The galaxies with spectroscopic redshifts are collected as the training sample to build a photo-z estimator and to assess the photo-z quality. As described in \citet{Zou2019}, we have compiled a spectroscopic redshift catalog from different spectroscopic surveys. Please refer to Table 2 in \citet{Zou2019} for the information and corresponding quality cuts. This redshift catalog is matched with the galaxy catalogs of DES and DESI using a matching radius of 1\arcsec. The number of matched galaxies are 469k and 2.8 million for DES and DESI, respectively. The HSC-SSP contains a value-added catalog of public spectroscopic redshifts \citep{Aihara2021}. This catalog supplements galaxies with higher redshift and fainter magnitudes, which can be more suitable for the photo-z estimation of the deeper HSC-SSP photometry. We select the galaxies in this value-added catalog with ``redshift $>$ 0 \& specz\_flag\_homogeneous = True" and obtain 636k galaxies with spectroscopic redshifts. Figure \ref{fig-zprop} shows the redshift and magnitude distributions of those spectroscopic galaxy samples in different surveys.

\section{Photo-z and stellar mass} \label{photoz}
\subsection{Photo-z estimation}
The photo-z estimation relies on the multi-wavelength photometric data that can construct spectroscopic energy distributions (SEDs) for galaxies. The methods to compute photo-z include template-fitting and machine-learning. The template-fitting method uses different types of modeled galaxy spectra to match the observed SED \citep{Bol2000,Ben2000,Brammer2008,Ilbert2009}. It is vital to construct proper theoretical spectral evolutionary models and to eliminate systematic bias in different photometric data when different survey data are combined. The machine-learning method tries to establish an empirical relation between the observed SED and redshift with a training sample that contains galaxies with known redshifts \citep{Carliles2010,Hogan2015,Sadeh2016}. This method is usually very efficient in both speed and accuracy, but it is usually difficult to build a farely representative training sample. This problem is greatly alleviated due to a large number of wide and deep extragalactic spectroscopic surveys.  

The photo-z estimation algorithm we adopt in this paper is similar to the local linear regression in \citep{Beck2016}, which has been used for the photo-z estimation with $ugriz$ photometry of the Sloan Digital Sky Survey (SDSS). We have applied this method to compute the photo-z with 7-band photometry of $ugrizW1W2$ by combining the SCUSS, SDSS, and WISE survey data \citep{Gao2018} and with 5-band photometry of $grzW1W2$ from the DESI and WISE \citep{Zou2019}. The local linear regression method assumes the relation between the photometric SED and redshift is linear in the local multi-dimensional color space. The locality of a galaxy is determined by the K Nearest Neighbor (KNN) algorithm, which selects $K$ galaxies in the training sample with shortest distances in color space. We use these $K$ nearest neighbors with known spectroscopic redshifts to derive the linear regression relation. This relation is then applied to the galaxy whose photo-z needs to be measured. During fitting the regression model, we use $3\sigma$ clipping algorithm to remove outliers. The root mean square error for the regression is considered as the photo-z uncertainty \citep{Beck2016,Gao2018}. The number of neighbours ($K$) is an important parameter to be determined. For a specified training set,  a too large value of $K$ might destroy the locality and increase the running rate of the photo-z algorithm. Conversely, a too-small value of $K$ might lead to inadequate neighbours to represent the locality in the color space and hence reduce the photo-z accuracy. We use a small subset of 10,000 galaxies in training sample to determine $K$. The photo-zs of these galaxies are estimated by the above method using a series of $K$ values ranging from 25 to 300 (interval of 25). The photo-z accuracy and outlier rate (see corresponding definitions in Section \ref{sec:pzquality}) are calculated and the best $K$ is chosen to make sure that the value is as small as possible and at the same time the photo-z accuracy and outlier rate approach to their lowest values. As a result, the selection of $K$ is different for different photometric datasets, which are 200 for DESI, 100 for DES, and 150 for HSP-SSP.


\subsection{Photo-z quality} \label{sec:pzquality}
The following quantities are defined to characterize the photo-z quality:
\begin{itemize}
\item \textbf{bias}: the systematic offset between the spectroscopic and photometric redshifts. The offset is defined as $\Delta z_\mathrm{norm} = (z_\mathrm{phot}-z_\mathrm{spec})/(1+z_\mathrm{spec})$, where {\zphot} is photo-z and {\zspec} is spectroscopic redshift.  The bias (\zbias) is calculated as the median value of {\dz} by iteratively applying 3$\sigma$ clipping algorithm to remove outliers.
\item \textbf{dispersion}: the dispersion (\zsigma) of {\dz}, which is also calculated by applying 3$\sigma$ clipping algorithm.  
\item \textbf{outlier rate}:  the fraction of galaxies with photo-zs deviating substantially from their spectroscopic redshifts. We adopt two definitions: one is {\pcon} that is defined as the fraction of galaxies with $|\Delta z_\mathrm{norm}| > 0.15$ and the other is {\pout} that is defined as the fraction of galaxies with $|\Delta z_\mathrm{norm}| > $3{\zsigma} after applying the 3$\sigma$ clipping algorithm. 
\end{itemize}

We present the photo-z quality in Table \ref{tab-surveyzquality}. The overall biases are ignorable and overall accuracies for DESI, DES, and HSC-SSP are 0.017, 0.024, and 0.029, respectively. The factors affecting the photo-z quality are complicated, which may include the adopted photometric system, photometric quality, galaxy samples, and spectroscopic training samples, etc.

\begin{table}
\setlength\tabcolsep{2pt}
\centering
\caption{Photo-z qualities for different imaging surveys \label{tab-surveyzquality}}
\begin{tabular}{c|c|c|c|c|c|c|c}
\hline
\hline
Survey & mag cut$^a$ & number$^b$ & redshift range$^c$ & Bias  &  Dispersion  & \pout & \pcon  \\
\hline
DESI  & $r<23$ & 320,060,206 & $z<1.0$ & 1.42e-4 & 1.72e-2&  6.26\% & 0.85\% \\
DES  & $r<24$ & 292,636,425  & $z<1.2 $ &7.36e-5 & 2.40e-2 & 8.02\% & 2.63\% \\
HSC-SSP & $r<25$  & 133,554,787 & $z<1.6$ & -2.72e-4 & 2.92e-2 & 10.02\% & 5.24\%  \\
\hline
\end{tabular}

Notes: $^a$Magnitude cut for selecting galaxies to determine the photo-z. $^b$Number of  galaxies selected using the  magnitude cut. $^c$General photo-z range. 
\end{table}

\begin{figure*}[htb!]
\center
\includegraphics[width=\linewidth]{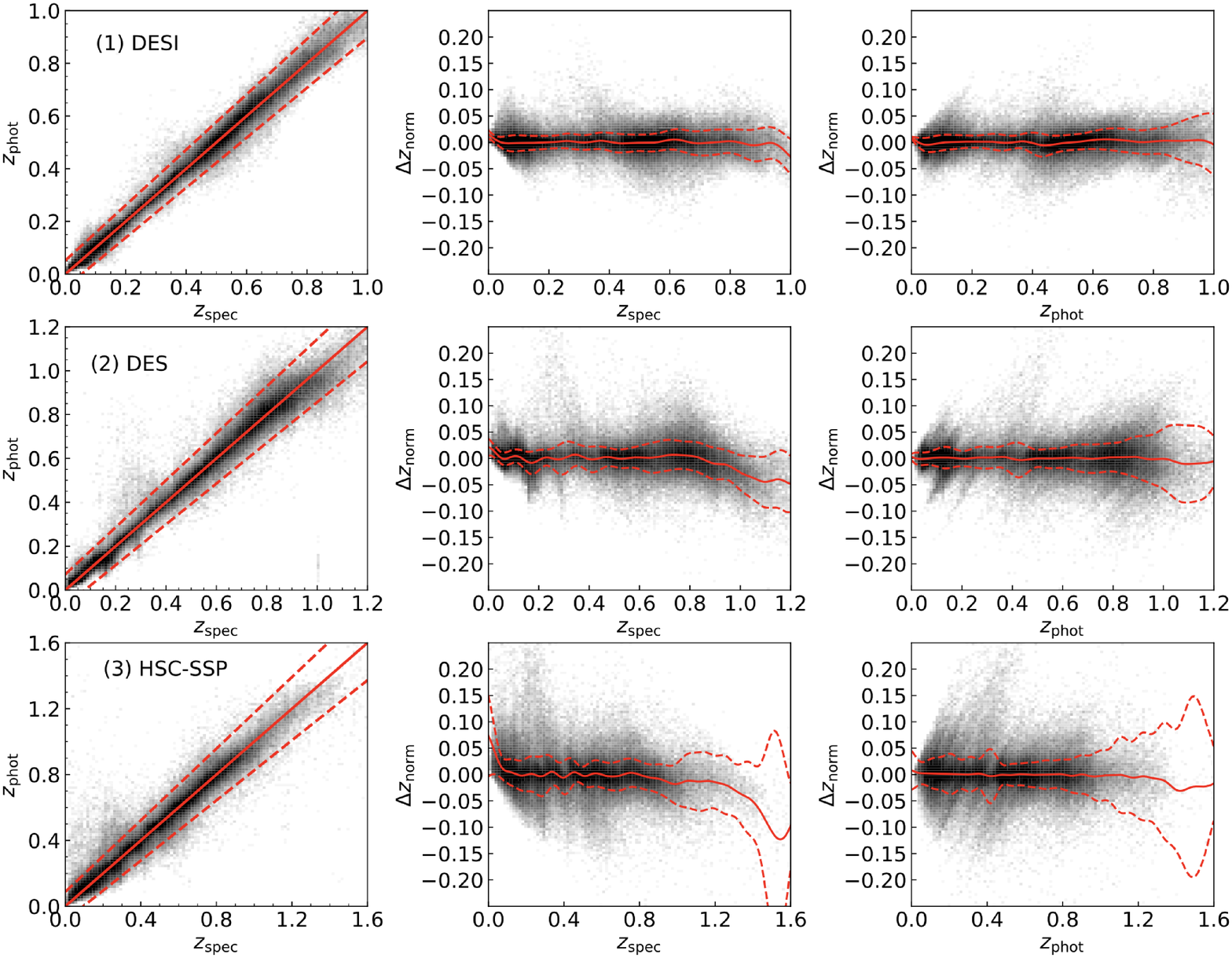}
\caption{Photo-z qualities for DESI (row 1), DES (row 2), and HSC-SSP (row 3). The left column presents the comparison between {\zphot} and {\zspec}. The solid line displays {\zphot  }$=$ {\zspec} and the dashed lines show $\Delta z_\mathrm{norm} \pm 3$\zsigma. The middle and right columns present {\zsigma} as function of {\zspec} and {\zphot}, respectively. The solid and dashed lines show the median and $1\sigma$ dispersion along the $x$ axis.   \label{fig-pzquality}}
\end{figure*}

Figure \ref{fig-pzquality} presents the comparisons between {\zphot} and {\zspec} and the photo-z accuracies as functions of {\zspec} and {\zphot} for the three survey datasets. Table \ref{tab-zzquality} presents the biases, dispersions, and outlier rates in different bins of {\zspec} and {\zphot}. The DESI photo-z accuracy is best, partly because the inclusion of WISE infrared photometry tends to select more red galaxies and depress the color-redshift degeneracy as discussed in \citet{Zou2019}.  The HSC-SSP data have a higher redshift coverage.  
	
\begin{center}
\begin{table}
\setlength\tabcolsep{2pt}
\centering
\small
\caption{Photo-z qualities in different redshift bins for different imaging surveys} \label{tab-zzquality}
\begin{tabular}{c|cccc|cccc|cccc}
\hline
\hline
 & \multicolumn{4}{c|}{DESI} & \multicolumn{4}{c|}{DES} & \multicolumn{4}{c}{HSC-SSP}  \\
\hline
{\zspec} & bias  &  dispersion  & {\pout} & {\pcon}  & bias  &  dispersion  & {\pout} & {\pcon}  & bias  &  dispersion  & {\pout} & {\pcon}    \\
\hline
(0.0,0.2) & -4.00e-04 & 1.37e-02 &  4.74\% &  0.53\% & 1.17e-03 & 1.54e-02 &  6.83\% &  1.71\% & 5.23e-03 & 2.68e-02 &  9.29\% &  5.76\%  \\
(0.2,0.4) & 1.42e-03 & 1.42e-02 & 10.20\% &  1.76\% & 1.47e-03 & 2.17e-02 & 13.57\% &  5.98\% & -2.98e-04 & 3.22e-02 &  8.63\% &  4.42\% \\
(0.4,0.6) & -1.53e-04 & 1.67e-02 &  5.01\% &  0.29\% & 7.99e-04 & 1.91e-02 &  8.22\% &  1.73\% & 3.55e-04 & 2.31e-02 &  8.67\% & 1.78\%  \\
(0.6,0.8) & 1.89e-03 & 2.17e-02 &  4.78\% &  0.53\%  & 4.21e-03 & 2.76e-02 &  5.94\% &  1.51\% & -9.89e-04 & 2.66e-02 &  8.26\% &  2.80\%  \\
(0.8,1.0) & -3.16e-03 & 2.83e-02 &  3.09\% &  1.06\% & -5.26e-03 & 3.01e-02 &  4.00\% &  1.25\% & -7.43e-03 & 2.90e-02 &  7.91\% &  3.32\%  \\
(1.0,1.2)  & & & & & -3.62e-02 & 4.39e-02 &  4.91\% &  5.83\% & -1.43e-02 & 5.01e-02 &  7.84\% &  8.42\%  \\
(1.2,1.4)  & & & &  & & & &  & -2.75e-02 & 5.38e-02 & 13.74\% & 17.07\% \\
(1.4,1.6) & & & &  & & & &   & -7.79e-02 & 9.98e-02 & 14.30\% & 35.74\%  \\
\hline
{\zphot} & bias  &  dispersion  & {\pout} & {\pcon}  & bias  &  dispersion  & {\pout} & {\pcon}  & bias  &  dispersion  & {\pout} & {\pcon}   \\
\hline
(0.0,0.2) & -7.65e-04 & 1.37e-02 &  3.96\% &  0.53\% & 6.07e-04 & 1.56e-02 &  5.14\% &  1.71\% & 1.79e-03 & 2.81e-02 &  3.71\% &  5.76\%  \\
(0.2,0.4) & 7.89e-04 & 1.39e-02 &  9.81\% &  1.76\% & 1.10e-03 & 2.05e-02 & 10.09\% &  5.98\% & 7.47e-04 & 3.21e-02 &  8.79\% &  4.42\%  \\
(0.4,0.6) & -8.80e-04 & 1.66e-02 &  6.96\% &  0.29\% & -4.26e-04 & 1.96e-02 & 11.37\% &  1.73\% & -9.28e-04 & 2.32e-02 & 13.02\% &  1.78\%  \\
(0.6,0.8) & 3.88e-03 & 2.05e-02 &  3.95\% &  0.53\% & 5.92e-04 & 2.56e-02 &  7.29\% &  1.51\% & -1.19e-04 & 2.51e-02 &  9.40\% &  2.80\%  \\
(0.8,1.0) & 3.12e-03 & 3.34e-02 &  5.79\% &  1.06\% & 1.24e-03 & 3.72e-02 &  4.97\% &  1.25\% & -1.51e-03 & 3.42e-02 &  9.71\% &  3.32\% \\
(1.0,1.2) & & & & & -7.01e-03 & 6.85e-02 &  4.77\% &  5.83\% & -1.28e-03 & 6.12e-02 & 13.33\% &  8.42\%  \\
(1.2,1.4)  & & & &  & & & &  & -8.52e-03 & 9.12e-02 & 12.49\% & 17.07\%  \\
(1.4,1.6)   & & & &  & & & & & -3.16e-02 & 1.00e-01 & 25.16\% & 35.74\%  \\
\hline
\end{tabular}
\end{table}
\end{center}

\begin{figure}
\center
\includegraphics[width=\linewidth]{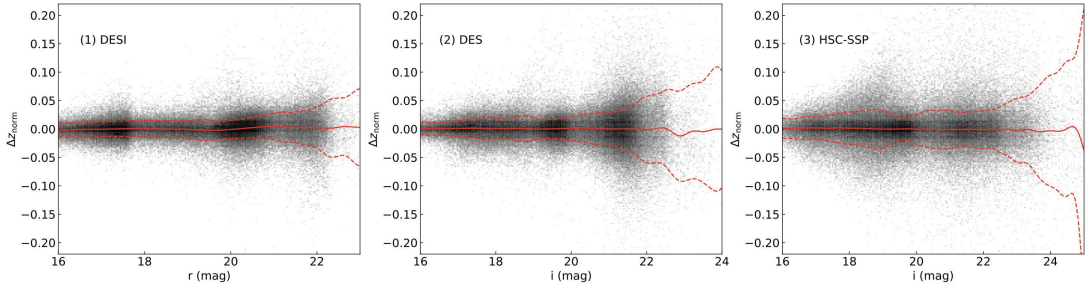}
\caption{Photo-z accuracy {\zsigma} as function of magnitude ($r$ band for DESI and $i$ band for others). The solid and dashed lines show the median and $1\sigma$ dispersion along the $x$ axis.   \label{fig-magzquality}}
\end{figure}

Figure \ref{fig-magzquality} shows the photo-z accuracy as function of magnitude and Table \ref{tab-magzquality} lists corresponding photo-z qualities in different magnitude ranges. As expected, both photo-z accuracy and outlier rate increase with magnitude. 

\begin{table}
\setlength\tabcolsep{2pt}
\centering
\small
\caption{Photo-z qualities in different magnitude bins for different imaging surveys} \label{tab-magzquality}
\begin{tabular}{c|cccc|cccc|cccc}
\hline
\hline
 & \multicolumn{4}{c|}{DESI} & \multicolumn{4}{c|}{DES} & \multicolumn{4}{c}{HSC-SSP}  \\
\hline
mag$^a$ & bias  &  dispersion  & {\pout} & {\pcon}  & bias  &  dispersion  & {\pout} & {\pcon}  & bias  &  dispersion  & {\pout} & {\pcon}   \\
\hline
(16,17) & -9.79e-04 & 1.27e-02 &  3.23\% &  0.00\% & 7.06e-04 & 1.18e-02 &  5.88\% &  0.07\% & 1.18e-03 & 1.89e-02 &  2.95\% &  0.03\% \\
(17,18) & 1.88e-04 & 1.34e-02 &  4.11\% &  0.04\% & 1.06e-03 & 1.71e-02 &  4.22\% &  0.32\% & 1.73e-03 & 2.42e-02 &  2.93\% &  0.11\%  \\
(18,19) & 2.28e-04 & 1.50e-02 &  4.44\% &  0.16\% & 4.22e-04 & 1.97e-02 &  5.50\% &  0.55\% & 6.54e-04 & 3.11e-02 &  4.75\% &  1.34\%  \\
(19,20) & -1.57e-03 & 1.69e-02 &  6.59\% &  0.49\% & 1.47e-04 & 1.76e-02 &  6.80\% &  1.41\% & -1.90e-04 & 2.37e-02 &  9.27\% &  2.91\%  \\
(20,21) & 1.76e-03 & 1.78e-02 &  5.49\% &  0.71\% & 4.68e-04 & 2.61e-02 &  7.67\% &  3.43\% & -9.06e-04 & 3.24e-02 & 11.68\% &  7.34\%  \\
(21,22) & 2.86e-03 & 2.78e-02 &  7.55\% &  3.03\% & 7.01e-05 & 3.68e-02 &  6.86\% &  4.37\% & -8.72e-04 & 3.11e-02 & 12.73\% &  8.06\%  \\
(22,23) & 3.43e-03 & 4.46e-02 &  5.88\% &  4.65\% & -3.57e-05 & 6.43e-02 &  5.67\% &  9.06\% & -1.44e-03 & 3.85e-02 & 11.73\% &  8.89\%  \\
(23,24) & & & &  & -3.41e-03 & 9.04e-02 &  5.97\% & 15.68\% & -1.48e-03 & 6.78e-02 & 12.46\% & 17.42\%  \\
(24,25)  & & & &  & & & & & -9.94e-04 & 1.21e-01 & 14.63\% & 31.58\%  \\
\hline
\end{tabular}

{Notes: $^a$the magnitude is referred to $r$ band for DESI and $i$ band for others.}
\end{table}

\subsection{stellar mass}
\begin{figure*}[htb!]
\center
\includegraphics[width=\linewidth]{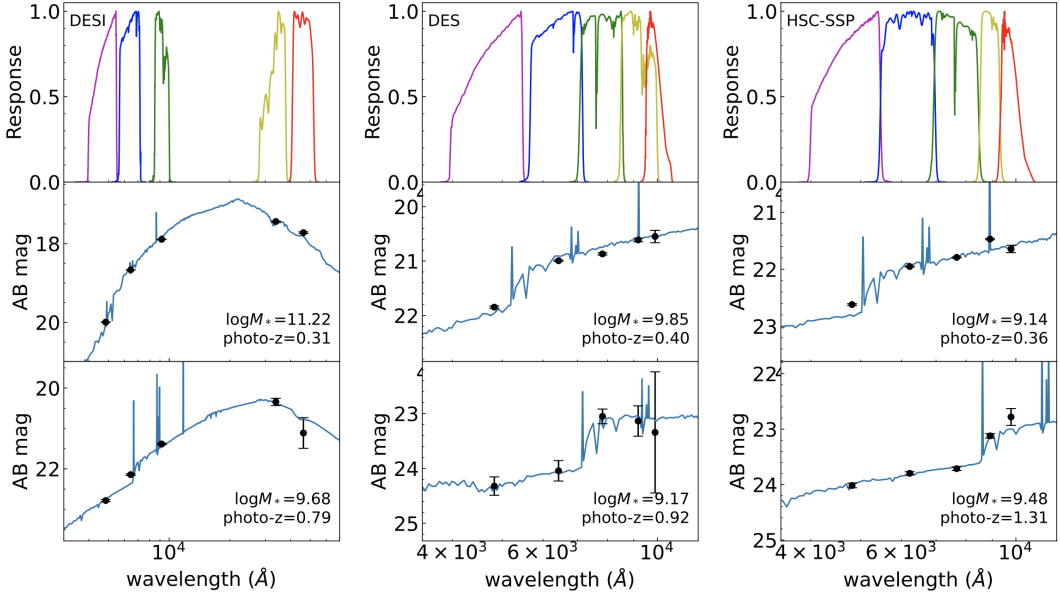}
\caption{Photometric filter set and two example SEDs for each of the three imaging surveys. The upper panels show the filter responses scaled to their maximums.  The lower and middle panels show two example SEDs, which are displayed in solid circles with error bars. The size of  the error bar presents the photometric error. The photo-z and stellar mass of each SED are marked in bottom-right corner of each panel. The best-fit template spectra are shown in dark-blue curves.  \label{fig-cspfit}}
\end{figure*}

Stellar mass is a fundamental physical quantity for galaxies. It can be derived by fitting the observed multi-wavelength spectral energy distribution (SED) with theoretical stellar population synthesis models. We adopt the LePhare software\footnote{\url{https://www.cfht.hawaii.edu/~arnouts/LEPHARE/lephare.html}} to estimate the stellar mass.  The redshift is fixed to photo-z as obtained in this paper. The default stellar population templates are used, which are constructed using the BC03 evolutionary models \citep{Bruzual2003} and \citet{Chabrier2003} initial mass function. These templates include the spectral models with 3 metallicities  (0.004, 0.008, and 0.02), 29 ages (0.01 Myr to 13.5 Gyr), and 9 exponentially declining star formation histories (timescale from 0.1 to 30 Gyr). Emission lines are added in the models. In addition, the model spectra are reddened using the extinction curve of \citet{Calzetti2000} and five $E(B-V)$ values of 0, 0.1, 0.2, 0.3, and 0.5 mag are adopted. The SED fitting provides both stellar mass and absolute magnitude in this paper. Figure \ref{fig-cspfit} shows some examples of the SED fitting. 

\begin{figure*}[htb!]
\center
\includegraphics[width=\linewidth]{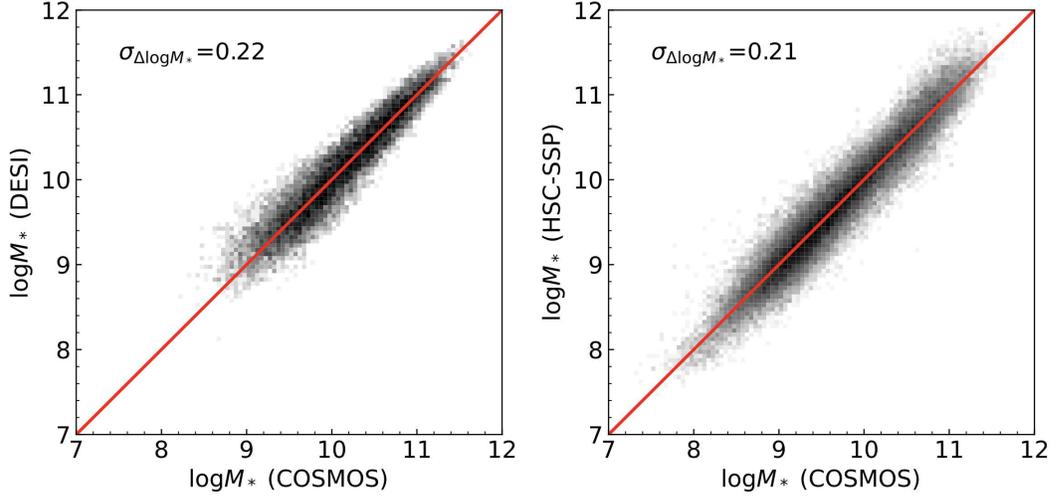}
\caption{Comparisons of the stellar mass in logarithm between our measurements and those from the COSMOS catalog (left for DESI and right for HSC-SSP). The dispersion of the mass difference ($\sigma_{\Delta\log M_*}$) is also displayed. \label{fig-masscomp}}
\end{figure*}

We compare our stellar mass ($M_*$) with that of \citet{Laigle2016}, who obtained accurate photo-zs and stellar masses for galaxies in the COSMOS field\footnote{\url{https://cosmos.astro.caltech.edu/}}. This field was covered by a total of 32 photometric bands ranging from ultraviolet to infrared, which ensures more convincing determination of stellar population properties from SED fitting. As adopted in this paper, the LePhare software was also used by \citet{Laigle2016}. To exclude the galaxies with large photo-z uncertainties and hence large uncertainties of stellar mass, we select galaxies with photo-z errors less than $0.1(1+z_\mathrm{phot})$ in our catalogs and less than $0.05(1+z_\mathrm{phot})$ in the COSMOS catalog. Figure \ref{fig-masscomp} shows the comparisons of the stellar mass. The general dispersion of the $\log M_*$ difference between our measurements and the COSMOS ones is about 0.2 dex. Although the COSMOS is out of the DES coverage,  we believe that the mass dispersion for DES should be at a similar level, because the photometric systems of DES and HSC-SSP are similar, the photo-z accuracies of these two surveys are at a similar level, and  the photometric depths are even better than DESI.

\section{Galaxy clusters} \label{sec:cluster} 
As the largest gravitationally bound systems in the universe, galaxy clusters have been effectively detected in large-scale optical surveys. There are two kinds of detection methods. One is based on the overdensity feature of the galaxy spatial distribution \citep{Szabo2011,Wen2012,Gao2020,Zou2021}. This detection method needs relatively accurate photo-z to probe the overdensities of galaxies above the average density of foreground and background galaxies. The other is based on the red-sequence feature of red galaxies, whose star formation has been quenched \citep{Koester2007,Hao2010,Rykoff2014}. This method recognizes the tight color distribution of red member galaxies in a cluster. It may lose some clusters without the red-sequence feature, which are very common at high redshift. We adopt a new fast cluster-finding algorithm as used in our previous papers to identify galaxy clusters for the DESI, DES, and HSC-SSP data. This cluster-finding method belongs to the detection methods based on the overdensity feature.

\subsection{Selection of galaxy sample}
In order to identify galaxy clusters, we only select the galaxies with relatively good photo-z and SED fitting: (1) the photo-z ranges are limited to $z_\mathrm{phot} <1.5$ for DESI and DES and $z_\mathrm{phot} <2$ for HSC-SSP; (2) the photo-z error is set to less than $0.1(1+z_\mathrm{phot})$; (3) the range of the $r$-band absolute magnitude is $-25 < M_r < -16$; (4) the stellar mass range is $6< \log M_* < 13$; (5) the logarithmic mass uncertainty is less than 0.4 dex. There are about 222, 221, and 101 million remaining galaxies for DESI, DES, and HSC-SSP, respectively. 

\subsection{Detecting clusters and assessing photo-z quality} 
The cluster detection method adopted here is a new clustering algorithm that can effectively find the overdensities of the galaxies over the sky. We give a brief introduction of this method as below. For more details, please refer to \citet{Zou2021}.
\begin{itemize}
\item[(1)] Galaxies in the photo-z catalog are subdivided into equal-area sky pixels in HEAPix format\footnote{\url{https://healpix.sourceforge.io/}}. The pixel area is about 0.84 deg$^2$. 
\item[(2)] The local density ($\rho$) of each galaxy in a sky pixel is calculated. It is defined as the number of galaxies with distance to this galaxy less than 0.5 Mpc and $\Delta z_\mathrm{norm} < 0.04$. When calculating the local density for a given galaxy, galaxies from the specified pixel and all its neighbour pixels are taken into account to avoid the boundary effect (area of about $9\times0.84=7.56$ deg$^2$).
\item[(3)] The background density of this galaxy ($\rho_\mathrm{bkg}$) is calculated in the above sky pixel and its neighbour pixels (total area is about 7.6 deg $^2$). It is the number of galaxies with distances to the specified galaxy larger than 1 Mpc and $\Delta z_\mathrm{norm} < 0.04$. 
\item[(4)] For each galaxy, a parameter $\theta$ is defined as the distance of the nearest galaxy with higher local density. 
\item[(5)] The density peaks ( or locations of galaxy clusters) are identified as the galaxies with large enough local density and distant enough away from other peaks, i.e, $\rho > n*\rho_\mathrm{bkg}$ and $\theta > 1$ Mpc, where $n$ is to be set. The brightest galaxy with distance to the peak smaller than 0.5 Mpc is considered as the brightest cluster galaxy (BCG, i.e. the cluster center).
\end{itemize}

\begin{figure}[htb!]
\center
\includegraphics[width=\linewidth]{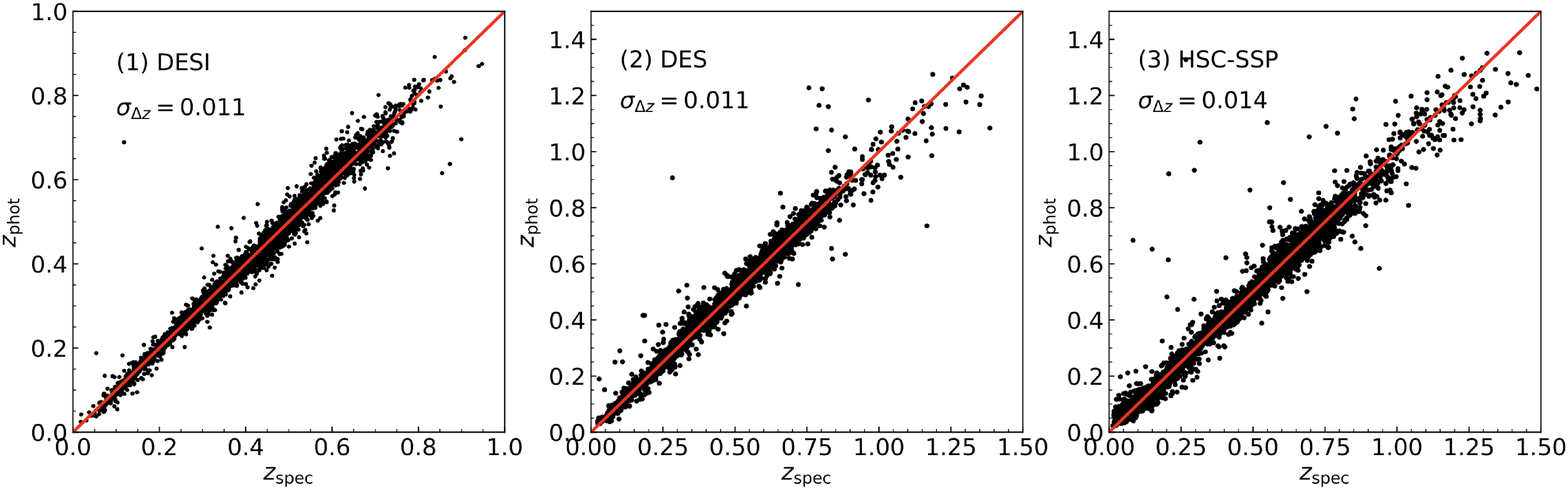}
\caption{Comparisons between {\zphot} and {\zspec} of BCGs for different surveys. The red line shows {\zphot} $=$ {\zspec}. The dispersion of {\zsigma} is displayed in each panel. \label{fig-clusterzquality}}
\end{figure}

Because the larger photo-z uncertainty for DES and HSC-SSP data suppresses the cluster overdensity relative to the background, a smaller threshold of $\rho$ is chosen. We set $n$ to 4 for DESI, 3.5 for DES, and 3 for HSC-SSP, which are roughly determined to assure relatively low false detection rates.  The above process can be easily executed in parallel mode. For each galaxy cluster, we calculate the number of member galaxies with distance to the center less than 1Mpc ($N_\mathrm{1MPC}$).  $N_\mathrm{1MPC}$ is subtracted the background density and is considered as a first-order estimate of the cluster richness. We only reserve relatively rich galaxy clusters with $N_\mathrm{1MPC} > 10$. The total numbers of detected galaxy clusters for DESI, DES, and HSC-SSP are 532,810, 86,963, and 36,566, respectively.  The number of clusters for DESI DR9 is somewhat smaller than that of DESI DR8 as presented in \citet{Zou2021}, which is partly due to slight different photometric data and selection of the galaxy samples. The photo-z accuracy of galaxy clusters is determined by comparing the {\zphot} and {\zspec} of BCGs. Figure \ref{fig-clusterzquality} shows these comparisons and displays the photo-z accuracies. Table \ref{tab-clusterzquality} summaries the photo-z qualities of galaxy clusters for the three datasets. 

\begin{table}[htb!]
\setlength\tabcolsep{3pt}
\centering
\caption{Photo-z qualities of galaxy clusters in different imaging surveys} \label{tab-clusterzquality}
\begin{tabular}{c|c|c|c|c|c|c}
\hline
\hline
Survey & N$^a$ & N$_\mathrm{spec}$$^b$ & Bias  &  Dispersion  & \pout  & F$^c$ \\
\hline
DESI  & 532,810 & 117,601 & 2.32e-5 & 1.08e-2&  2.85\% & 6.98\% \\
DES  &  86,963 & 5,035 &6.71e-4 & 1.10e-2 & 4.79\%  & 8.68\% \\
HSC-SSP & 36,566 & 6,267 & 2.40e-5 & 1.44e-2 & 4.56\%  &5.65\% \\
\hline
\end{tabular}

{Notes:  $^a$Number of galaxy clusters. $^b$Number of galaxy clusters having spectroscopic redshifts. $^c$False detection rate.}
\end{table}

We follow the same process as described in \citet{Zou2021} to estimate the false detection rate of our cluster-finding method. A Monte Carlo simulation based on the actual photometric data is performed to generate a mock catalog: (1) galaxies are redistributed by randomly moving away from their original positions within the distance of 1--2.5 Mpc; (2) the properties including redshift of galaxies are shuffled. The shuffled galaxies could be regarded as a random redistribution of their original positions in the 3-D space  and meanwhile maintain the correlated large-scale structure to some degree. In this way, the overdensity of galaxy clusters should be shuffled out. Then we apply the same detecting method as used in this paper  to the mock catalog to assess the false detection. We should note that this kind of simulation might underestimate the false detection, because the projection effects include the impacts from both correlated and uncorrelated large-scale structures. The false detection rate $F$ is defined as the ratio of the number of clusters detected in the mock catalog to that of the original catalog. The last column of Table \ref{tab-clusterzquality} lists the false rate for each survey. 

\subsection{Total mass}
\begin{figure*}[htb!]
\center
\includegraphics[width=\linewidth]{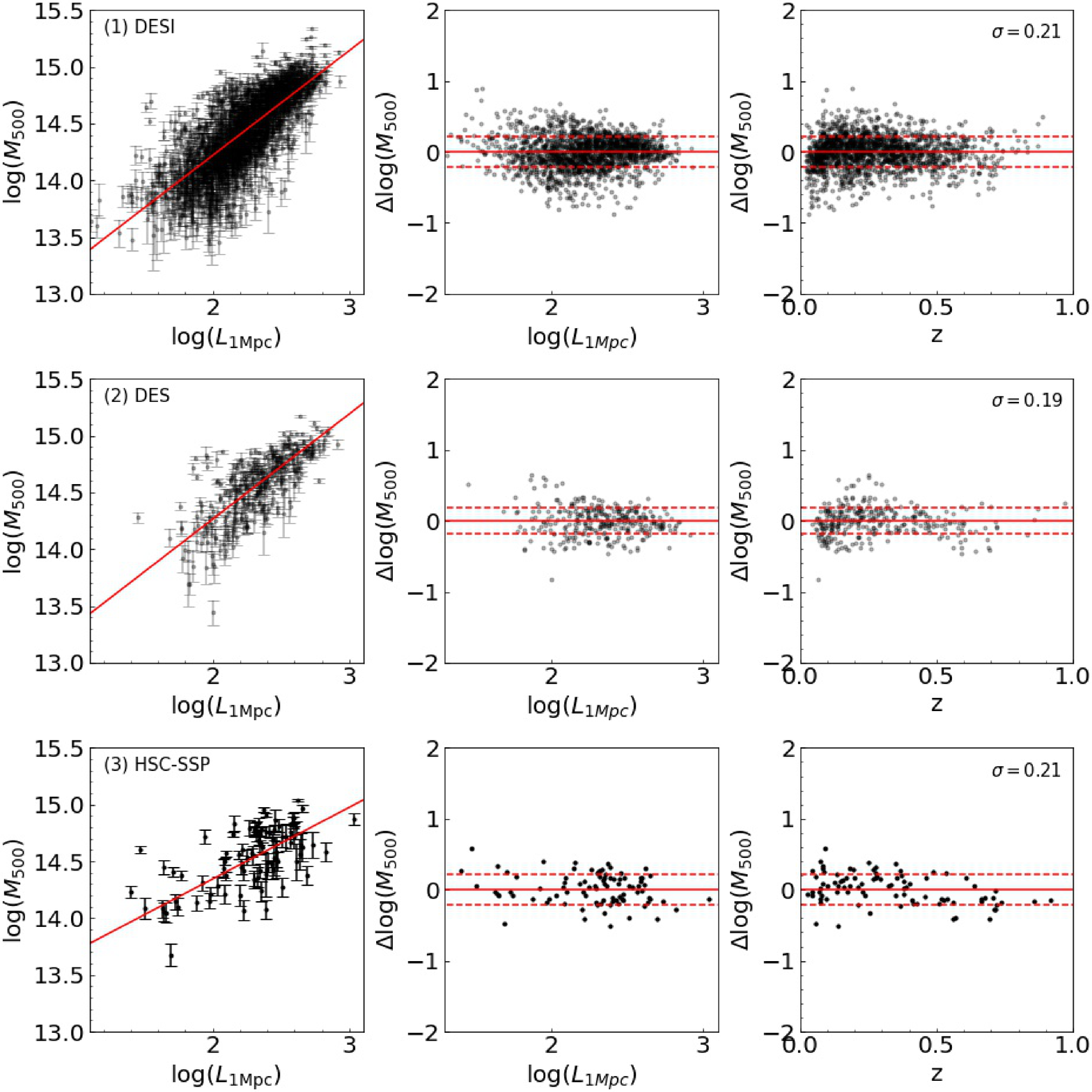}
\caption{The left column shows the logarithmic calibration relations between $M_{500}$ and $L_\mathrm{1Mpc}$ for DESI (row 1), DES (row 2), and HSC-SSP (row 3). The middle and right columns show $\Delta \log M_{500}$ as functions of $\log L_\mathrm{1Mpc}$ and redshift, respectively. Here $\Delta\log M_{500}$ is the difference between the measurements and the linear predictions.  The red solid lines display $\Delta\log M_{500} = 0$ and the red dashed line show the $1\sigma$ dispersion of $\Delta\log M_{500}$, which is also marked on the rightest panel. \label{fig-calibration}}
\end{figure*}

The total mass of galaxy clusters (including baryon and dark matters) can be effectively estimated from the measurements of weak gravitational lensing or observations of X-ray emission and Sunyaev \& Zel'dovich (SZ) effect in microwave band. We have compiled a catalog of 3,157 galaxy clusters with the total mass ($M_{500}$) estimated using the X-ray and SZ observations \citep{Zou2021}. It can be used for calibrating the mass of our detected clusters. The optical luminosity of member galaxies in a cluster is a good proxy of the cluster richness and hence can be used to estimate the total mass. We define $L_\mathrm{1Mpc}$ as the total $r$-band luminosity of member galaxies. $L_\mathrm{1Mpc}$ is also subtracted the background luminosity, which is calculated in the same way as $N_\mathrm{1Mpc}$.  We find that the richness $L_\mathrm{1Mpc}$ presents a good linear relation with the total mass in the logarithmic space  and this relation is independent on the redshift (see Figure \ref{fig-calibration}). The calibration relation is described as $\log(M_{500}) = a\log(L_\mathrm{1Mpc})+b$, where $a$ and $b$ are coefficients to be fitted. We derive these calibration relations for different surveys and apply them to our detected clusters. The overall calibration accuracy is about 0.2 dex. Here we assume that the above linear calibration relations are applicable for the clusters with richness and redshift out of the coverage of the calibration catalog. Note that the calibrations might suffer a little from the Malmquist bias as the cluster sample is constructed with flux-limited X-ray and SZ observations. Table \ref{tab-calibration} lists the calibration coefficients for different survey data. The characteristic radius $R_{500}$ is calculated from the relation of $M_{500} = \frac{4\pi}{3}R_{500}^3\times500\rho_c$, where $\rho_c$ is the critical density of the universe.

\begin{center}
\begin{table}[htb!]
\centering
\caption{Calibrations of the total mass for different surveys} \label{tab-calibration}
\begin{tabular}{c|c|c|c|c}
\hline
\hline
Survey & N$^a$ & $a$  &  $b$  &  $\sigma_{\Delta\log M_{500}}$  \\
\hline
DESI  & 1747 & 0.92$\pm$0.06 & 11.41$\pm$0.80 & 0.21  \\
DES  &  310 & 0.93$\pm$0.06 & 12.41$\pm$0.21 & 0.19  \\
HSC-SSP & 88 & 0.63$\pm$ 0.09 & 13.08$\pm$0.20 & 0.21   \\
\hline
\end{tabular}
\end{table}
\end{center}

\begin{figure*}[htb!]
\center
\includegraphics[width=\linewidth]{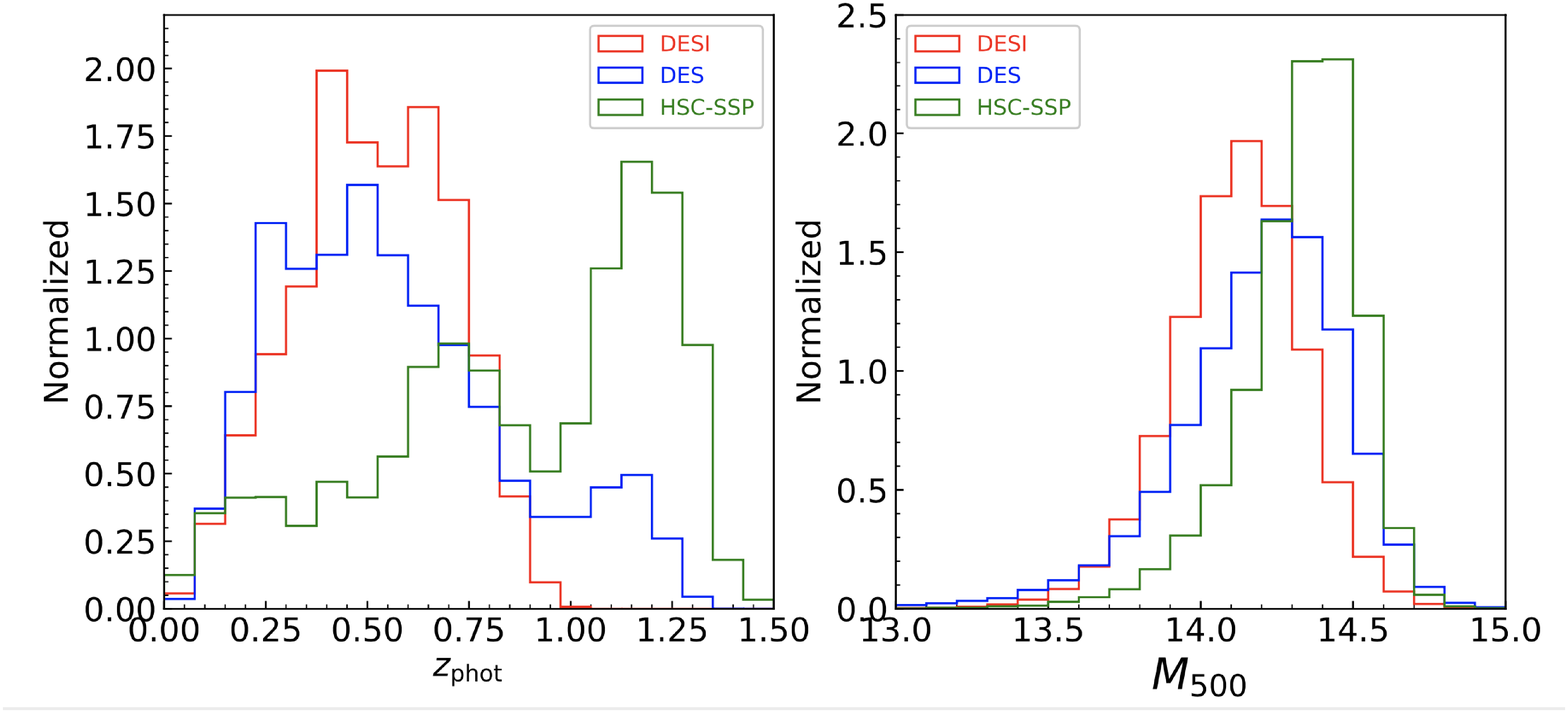}
\caption{Left: normalized redshift distribution of our clusters. Right: normalized mass distribution of our clusters. \label{fig-zmass}}
\end{figure*}

Figure \ref{fig-zmass} presents the redshift and mass distributions of our clusters. The  median redshifts for DESI, DES, and HSC-SSP are 0.52, 0.52, and 0.96, respectively. The median logarithmic masses for DESI, DES, and HSC-SSP are 14.13, 14.23, and 14.36, respectively. The HSC-SSP survey is deeper and can extend to higher redshift, so the average total mass of HSC-SSP clusters should be larger to be detected.

\subsection{Comparison with redMaPPer clusters}
As a representative cluster-finding method based on the red-sequence feature, redMaPPer has been designed to identify galaxy clusters in a few large-scale photometric surveys \citep{Rykoff2014,Rykoff2016}. The redMaPPer cluster catalog for the SDSS DR8 is used for comparison with our catalogs. The latest version of v6.3\footnote{\url{http://risa.stanford.edu/redmapper/}} is obtained, which includes 26,111 clusters and covers the redshift range of $0.08 < z < 0.55$. The photometric redshift uncertainty of the redMaPPer clusters is at the level of \zsigma$\sim0.01$. There are 25840, 1979, and 2222 redMaPPer clusters covered by DESI, DES and HSC-SSP, respectively. 

\begin{figure*}[htb!]
\center
\includegraphics[width=\linewidth]{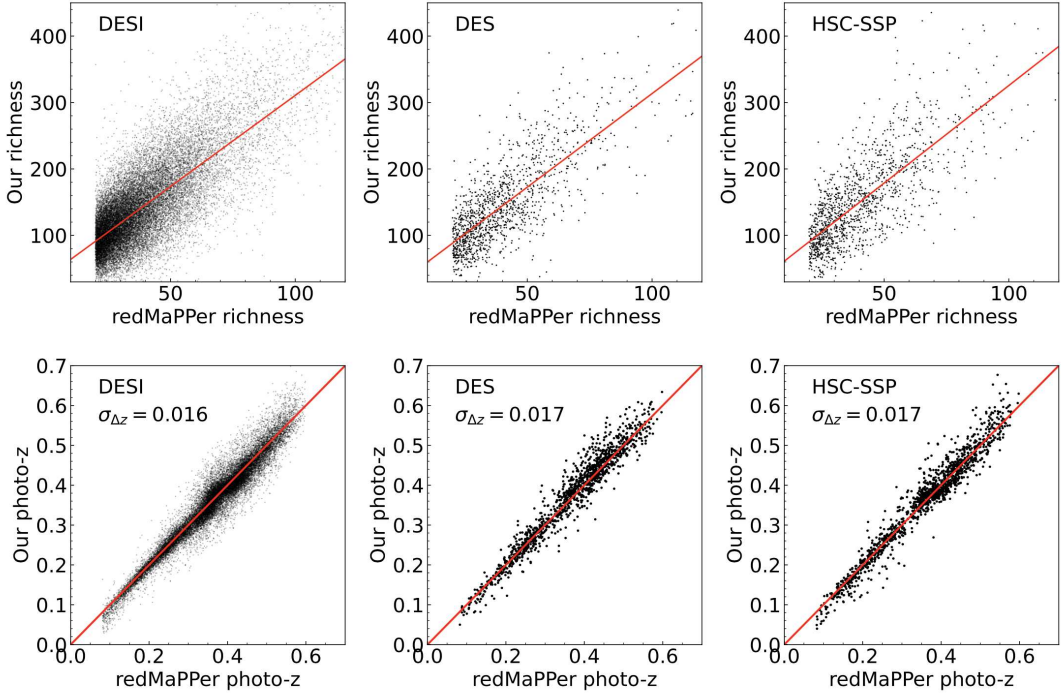}
\caption{Richness (upper panels) and redshift (lower panels) comparisons of our detected clusters with the SDSS DR8 redMaPPer cluster catalog. The red line in the upper panels is the best linear fit of the relation between two richness measurements. The red line in the lower panels presents $y=x$. The dispersions ({\zsigma}) are displayed in these panels, where photo-z from the redMaPPer is regarded as {\zspec}.  \label{fig-clustercomp}}
\end{figure*}

We match our catalogs with the redMaPPer catalog using a redshift tolerance of \dz$ < 0.06$ and a projection separation of 1 Mpc. The number of matched clusters are 25096 (97.1\%), 1319 (66.6\%), and 1338 (60.2\%) for DESI, DES, and HSC-SSP, respectively.  The relatively lower matching rates for DES and HSC-SSP are mainly due to larger photo-z  uncertainties, which smooth out some of low-level overdensities. Figure \ref{fig-clustercomp} shows the comparisons of richness and photo-z between our catalogs and redMaPPer catalog. The richnesses of matched clusters in these catalogs presents good correlations. We obtain the following relations by linear fitting:
\begin{eqnarray}
R_\mathrm{DESI} &=& 2.74R_\mathrm{redMaPPer}+36.45,  \nonumber \\ 
R_\mathrm{DES} &=& 2.82R_\mathrm{redMaPPer}+31.77,  \nonumber \\
R_\mathrm{HSC-SSP} &=& 2.94R_\mathrm{redMaPPer}+31.81,  \nonumber 
\end{eqnarray}
where $R$ is the richness. We can also see from Figure \ref{fig-clustercomp} that the photo-zs of our cluster catalogs present excellent consistency with that of the redMaPPer catalog. The general dispersion of {\dz} is about 0.017. 

\section{summary} \label{sec:summary}
As more and more wide and deep imaging surveys have been carried out, the photo-z technique becomes critical to their scientific achievements. Photo-z can be effectively derived from multi-wavelength photometric observations and it is a basic parameter to infer other physical properties of galaxies and to explore the galaxy evolution, especially in the early universe, where spectroscopic observations are difficult. 

Recently, several large-scale wide and deep imaging surveys, including DESI, DES, and HSC-SSP, have released their latest data. For certain reasons, DES and HSC-SSP have not published photo-z measurements. We have successfully applied a local linear regression algorithm to estimate the photo-z to the SCUSS and DESI DR8 data. In this paper, we apply the same method to derive the photo-z for galaxies in the latest data of the above three imaging surveys.  With spectroscopic training data, we construct a linear regression model for each galaxy in the local color space and estimate the photo-z using this model. The photo-z uncertainties for DESI, DES, and HSC-SSP is about 0.017, 0.024, and 0.029, respectively. In addition to photo-z, a series of stellar population properties of galaxies including the stellar mass are derived by using the SED fitting method. The redshifts of galaxies are fixed to photo-z we derived in this paper. The uncertainty of logarithmic stellar mass is about 0.2 dex.

With the photo-z catalogs, we try to detect galaxy clusters using a fast cluster-finding method, which was also successfully applied to the SCUSS and DESI DR8 data. Galaxy clusters are considered as the overdensties with large-enough local galaxy densities and substantial separations from each other. The number of detected galaxy clusters with members larger than 10 are 532,810, 86,963, and 36,566 for DESI DR9, DES, and HSC-SSP, respectively. The number of galaxy clusters we detected is by far the largest. Monte-Carlo simulations present the false detection rate is about  6--8\%. The photo-z accuracy for our galaxy clusters is about 0.011--0.014. Both redshift and richness show good consistency with those of the well-known redMaPPer clusters.

The catalogs we construct in this paper will be made publicly accessible at the Science Data Bank (ScienceDB\footnote{\url{https://www.doi.org/10.11922/sciencedb.o00069.00003}}) and PaperData Repository\footnote{\url{https://doi.org/10.12149/101089}}. The series of work on photo-z and galaxy clusters we conduct can be extended to future imaging surveys using the Chinese space-based and ground-based facilities, such as the China Space Station Telescope \citep[CSST;][]{Zhan2021}, SiTian Project \citep{Liu2021}, Multi-channel Photometric Survey Telescope (Mephisto\footnote{\url{http://www.swifar.ynu.edu.cn/info/1015/1073.htm}}), and Wide Field Survey Telescope (WFST\footnote{\url{http://wfst.ustc.edu.cn/}}). 

Here we provide some general guidelines and notes for users to use the photo-z and cluster catalogs. In the future studies with these catalogs, we will have further investigations of all possible issues that are not fully analyzed in this paper. 
\begin{itemize}
\item[(1)]  From the photo-z statistics,  it seems that the HSC-SSP and  DES photo-z accuracies are worse than the DESI one. Actually, it is not quite true. As mentioned before,  the inclusion of WISE photometry tends to select redder galaxies whose colors are more sensitive to redshift. If we select the same galaxies for comparison, the photo-z accuracies for these three surveys are similar.  However, the HSC-SSP  and DES data are much deeper than DESI. If the users need the photo-zs and corresponding stellar population properties of fainter and more distant galaxies, the HSC-SSP and DES data are better choices. 
\item[(2)] The photo-z qualities are depended on the properties of galaxies.  For example, we have already known that at the same redshift, the photo-z quality for blue galaxies might be twice worse than red galaxies. Before using the photo-z catalog, we suggest that users could first apply the spec-zs in our catalog to assess the quality of photo-z in the color space and select specific galaxies they want.
\item[(3)] The users should notice that we only select morphologically-classified galaxies. It may lose some point-like galaxies, which might be faint or distant. In addition, the magnitude cuts of $r$ and $i$ bands may also lead to miss some high-redshift galaxies. Actually, we lack the training samples of faint and distant galaxies to reliably estimate their photo-zs. Future spectroscopic surveys would improve this situation.
\item[(4)] For stellar population properties in the photo-z catalog, the stellar mass is fully tested and should be most reliable. However, the users may be cautious in using other parameters such as star formation rate and stellar age. 
\item[(5)] Compared with the DESI clusters, the number of galaxy clusters we identified from DES and HSC-SSP surveys should be somewhat  underestimated. It is because that the worse photo-z accuracy smooths the overdensity feature in the three-dimensional space and leads to less detections of low-richness clusters. 
\item[(6)] The BCG is identified as the brightest galaxy in a specified redshift slice with the distance from the density peak less than 0.5 Mpc. Whereas, the relatively large photo-z uncertainty may cause a wrong identification. We have visually checked several hundreds of rich clusters and found that the false rate is quite low. More quantitative analyses will be conducted in future. 
\end{itemize}

\begin{acknowledgements}
We thank the anonymous referee for his/her thoughtful comments and insightful suggestions that improve our paper greatly. This work is supported by the National Natural Science Foundation of China (NSFC) under grant 12120101003 and Beijing Municipal Natural Science Foundation under grant 1222028. We acknowledge the science research grants from the China Manned Space Project with Nos. CMS-CSST- 2021-A02 and CMS-CSST-2021-A04. The work is also supported by NSFC under grants 11890691, 11890693, 11873053, 12073035, 12133010, 11733007, and the National Key R\&D Program of China under grant 2019YFA0405501. 

The Legacy Surveys consist of three individual and complementary projects: the Dark Energy Camera Legacy Survey (DECaLS; NOAO Proposal ID \# 2014B-0404; PIs: David Schlegel and Arjun Dey), the Beijing-Arizona Sky Survey (BASS; NOAO Proposal ID \# 2015A-0801; PIs: Zhou Xu and Xiaohui Fan), and the Mayall z-band Legacy Survey (MzLS; NOAO Proposal ID \# 2016A-0453; PI: Arjun Dey). DECaLS, BASS and MzLS together include data obtained, respectively, at the Blanco telescope, Cerro Tololo Inter-American Observatory, National Optical Astronomy Observatory (NOAO); the Bok telescope, Steward Observatory, University of Arizona; and the Mayall telescope, Kitt Peak National Observatory, NOAO. The Legacy Surveys project is honored to be permitted to conduct astronomical research on Iolkam Du'ag (Kitt Peak), a mountain with particular significance to the Tohono O'odham Nation.

This project used public archival data from the Dark Energy Survey (DES). Funding for the DES Projects has been provided by the U.S. Department of Energy, the U.S. National Science Foundation, the Ministry of Science and Education of Spain, the Science and Technology FacilitiesCouncil of the United Kingdom, the Higher Education Funding Council for England, the National Center for Supercomputing Applications at the University of Illinois at Urbana-Champaign, the Kavli Institute of Cosmological Physics at the University of Chicago, the Center for Cosmology and Astro-Particle Physics at the Ohio State University, the Mitchell Institute for Fundamental Physics and Astronomy at Texas A\&M University, Financiadora de Estudos e Projetos, Funda{\c c}{\~a}o Carlos Chagas Filho de Amparo {\`a} Pesquisa do Estado do Rio de Janeiro, Conselho Nacional de Desenvolvimento Cient{\'i}fico e Tecnol{\'o}gico and the Minist{\'e}rio da Ci{\^e}ncia, Tecnologia e Inova{\c c}{\~a}o, the Deutsche Forschungsgemeinschaft, and the Collaborating Institutions in the Dark Energy Survey.

The Hyper Suprime-Cam (HSC) collaboration includes the astronomical communities of Japan and Taiwan, and Princeton University. The HSC instrumentation and software were developed by the National Astronomical Observatory of Japan (NAOJ), the Kavli Institute for the Physics and Mathematics of the Universe (Kavli IPMU), the University of Tokyo, the High Energy Accelerator Research Organization (KEK), the Academia Sinica Institute for Astronomy and Astrophysics in Taiwan (ASIAA), and Princeton University. Funding was contributed by the FIRST program from the Japanese Cabinet Office, the Ministry of Education, Culture, Sports, Science and Technology (MEXT), the Japan Society for the Promotion of Science (JSPS), Japan Science and Technology Agency (JST), the Toray Science Foundation, NAOJ, Kavli IPMU, KEK, ASIAA, and Princeton University. 

This paper is based on data collected at the Subaru Telescope and retrieved from the HSC data archive system, which is operated by the Subaru Telescope and Astronomy Data Center (ADC) at NAOJ. Data analysis was in part carried out with the cooperation of Center for Computational Astrophysics (CfCA), NAOJ. We are honored and grateful for the opportunity of observing the Universe from Maunakea, which has the cultural, historical and natural significance in Hawaii. 

\end{acknowledgements}

\appendix                 

\section{Data Availability}

The photo-z and cluster catalogs in this paper are available at ScienceDB. The download link is \textit{\url{https://www.doi.org/10.11922/sciencedb.o00069.00003}}. The PaperData Repository provides a backup accessing address (\textit{\url{https://doi.org/10.12149/101089}}). The corresponding structure of the data storage is shown as below: 
\begin{itemize}
\item[-] \textit{desdr2\_galaxy\_cspcat.fits} (file): the total catalog of photo-z and stellar mass for DES;
\item[-] \textit{desidr9\_galaxy\_cspcat.fits} (file): the total catalog of photo-z and stellar mass for DESI;
\item[-] \textit{hscpdr3\_wide\_galaxy\_cspcat.fits} (file): the total catalog of photo-z and stellar mass for HSC-SSP;
\item[-] \textit{photoz\_desdr2} (directory): dividing \textit{desdr2\_galaxy\_cspcat.fits} into small files with file names of \textit{desdr2\_galaxy\_cspcat\_raXXX\_YYY.fits}, where XXX and YYY are the lower and upper limits of R.A in degrees, respectively;
\item[-] \textit{photoz\_desidr9} (directory): dividing \textit{desidr9\_galaxy\_cspcat.fits} into small files with file names of \textit{desidr9\_galaxy\_cspcat\_raXXX\_YYY.fits};
\item[-] \textit{photoz\_hscpdr3} (directory): dividing \textit{hscpdr3\_wide\_galaxy\_cspcat.fits} into small files with file names of \textit{hscpdr3\_wide\_galaxy\_cspcat\_raXXX\_YYY.fits};
\item[-] \textit{galaxy\_clusters\_desdr2.fits} (file): the catalog of galaxy clusters for DES;
\item[-] \textit{galaxy\_clusters\_desidr9.fits} (file): the catalog of galaxy clusters for DESI;
\item[-] \textit{galaxy\_clusters\_hscpdr3\_wide.fits} (file): the catalog of galaxy clusters for HSC-SSP;
\item[-] \textit{readme.txt} (file): a brief instruction of the data.
\end{itemize}

\section{The photo-z catalogs}
Table \ref{tab:desi_pz}, \ref{tab:des_pz}, and \ref{tab:hsc_pz} present the content of the photo-z catalogs for DESI, DES, and HSC-SSP, respectively. Note that the stellar population parameters other than stellar mass and absolute magnitude are not fully tested, but they are kept in the catalog in case users find that they are useful.

\begin{center}
\begin{longtable}{lll}
\caption{Column description of the photo-z catalog for DESI} \label{tab:desi_pz}\\
 \hline
  \hline
 Column & Unit  & Description \\
  \hline
  \endfirsthead
 \multicolumn{3}{c}%
{\tablename\ \thetable\ -- \textit{Continued from previous page}} \\
\hline
 \hline
Column & Unit  & Description \\
\hline
\endhead
\hline \multicolumn{3}{r}{\textit{Continued on next page}} \\
\endfoot
\hline
\endlastfoot

 ID &  & Unique object ID  \\
 RA  & degree &  R.A. in J2000  \\
 DEC  & degree & Declination in J2000 \\
 MAG\_G  & mag  &  $g$-band magnitude \\
 MAG\_R  & mag   &  $r$-band magnitude \\
 MAG\_Z  & mag    &  $z$-band magnitude \\
 MAG\_W1  & mag &   $W1$-band magnitude \\
 MAG\_W2  & mag  &  $W2$-band magnitude \\
 MAGERR\_G  & mag  &  $g$-band magnitude error \\
 MAGERR\_R  & mag  &  $r$-band magnitude error \\
 MAGERR\_Z  & mag  &  $z$-band magnitude error \\
 MAGERR\_W1  & mag  & $W1$-band magnitude error \\
 MAGERR\_W2  & mag  & $W2$-band magnitude error \\
 GALDEPTH\_G  & mag &  $5\sigma$ galaxy depth in $g$ band \\
 GALDEPTH\_R & mag &  $5\sigma$ galaxy depth in $r$ band \\
 GALDEPTH\_Z  & mag &  $5\sigma$ galaxy depth in $z$ band \\
 TYPE & & Morphological type \\
 SHAPE\_R & arcsec & Half-light radius of galaxy model for galaxy type type \\
 SHAPE\_R\_IVAR & 1/arcsec$^2$ & Inverse variance of SHAPE\_R \\
 SHAPE\_E1 &  & Ellipticity component 1 of galaxy model \\
 SHAPE\_E1\_IVAR &  & Inverse variance of SHAPE\_E1 \\
 SHAPE\_E2 &  & Ellipticity component 2 of galaxy model  \\
 SHAPE\_E2\_IVAR &  & Inverse variance of SHAPE\_E2 \\
 SERSIC &  & Power-law index for the Sersic profile model \\
 SERSIC\_IVAR &  & Inverse variance of SERSIC \\
 N\_NEIGHBOUR & & Number of neighbours used for photo-z estimation \\
 N\_FILTER   & &  Number of filters used \\
 PHOTO\_Z   & &  Estimated photometric redshift \\
 PHOTO\_ZERR & & Estimated photometric redshift error \\
 SPEC\_Z  & &    Spectroscopic redshift if available \\
 MEAN\_Z   & &   Mean spectroscopic redshift of $k$ nearest neighbours \\
 SIGMA\_Z   & &   Standard deviation of the spectroscopic redshifts of $k$ nearest neighbours \\
 NEAREST\_Z  & &  Spectroscopic redshift of the nearest neighbour \\
 MEAN\_DIS  & &  Mean Euclidean distance in the color space of neighbours \\
 CHI\_BEST  & &   $\chi^2$ of the best BC03 model fitting \\
 MOD\_BEST & & ID of the best-fitted BC03 model \\
 EBV\_BEST & mag &  Best estimated E(B-V) in three values of 0, 0.5, and 1.0 mag \\
 SCALE\_BEST & & Scaling factor between observed SED and model SED \\
 MAG\_ABS\_G & mag & $g$-band absolute magnitude \\
 MAG\_ABS\_R & mag & $r$-band absolute magnitude \\
 MAG\_ABS\_Z & mag & $z$-band absolute magnitude \\
 MAG\_ABS\_W1 & mag & $W1$-band absolute magnitude \\
 MAG\_ABS\_W2 & mag & $W2$-band absolute magnitude \\
 AGE\_BEST & yr &   Best estimated age \\
 AGE\_INF & yr &    Lower limit of age with 68\% confidence level \\
 AGE\_MED  & yr &   Median age \\
 AGE\_SUP & yr &    Upper limit of age with 68\% confidence level \\
 MASS\_BEST & dex &  Best estimated log stellar mass ($M_{\sun}$) \\
 MASS\_INF  & dex &  Lower limit of log stellar mass with 68\% confidence level \\
 MASS\_MED & dex &   Median log stellar mass \\
 MASS\_SUP  & dex &  Upper limit of log stellar mass with 68\% confidence level \\
 SFR\_BEST & $M_{\sun}$ yr$^{-1}$ &   Best estimated SFR \\
 SFR\_INF   & $M_{\sun}$ yr$^{-1}$ &  Lower limit of SFR with 68\% confidence level \\
 SFR\_MED & $M_{\sun}$ yr$^{-1}$ &  Median SFR \\
 SFR\_SUP  & $M_{\sun}$ yr$^{-1}$ &  Upper limit of SFR with 68\% confidence level \\
 SSFR\_BEST & yr$^{-1}$ &  Best estimated SSFR \\
 SSFR\_INF & yr$^{-1}$ & Lower limit of SSFR with 68\%confidence level \\
 SSFR\_MED  & yr$^{-1}$ &   Median SSFR \\
 SSFR\_SUP  & yr$^{-1}$ &   Upper limit of SSFR with 68\%confidence level \\
 LUM\_NUV\_BEST & dex &  Best estimated log NUV luminosity ($L_{\sun}$) \\
 LUM\_R\_BEST & dex &   Best estimated log R-band luminosity ($L_{\sun}$) \\
 LUM\_K\_BEST & dex &  Best estimated log K-band luminosity ($L_{\sun}$) \\
\end{longtable}
\end{center}

\footnotesize
\begin{center}
\begin{longtable}{lll}
\caption{Column description of the photo-z catalog for DES} \label{tab:des_pz}\\
 \hline
  \hline
 Column & Unit  & Description \\
  \hline
  \endfirsthead
 \multicolumn{3}{c}%
{\tablename\ \thetable\ -- \textit{Continued from previous page}} \\
\hline
 \hline
Column & Unit  & Description \\
\hline
\endhead
\hline \multicolumn{3}{r}{\textit{Continued on next page}} \\
\endfoot
\hline
\endlastfoot
 COADD\_OBJECT\_ID &  & Unique object ID in DES \\
 RA  & degree &  R.A. in J2000  \\
 DEC  & degree & Declination in J2000 \\
 MAG\_G  & mag  &  $g$-band magnitude \\
 MAG\_R  & mag   &  $r$-band magnitude \\
 MAG\_I  & mag    &  $i$-band magnitude \\
 MAG\_Z  & mag &   $z$-band magnitude \\
 MAG\_Y  & mag  &  $Y$-band magnitude \\
 MAGERR\_G  & mag  &  $g$-band magnitude error \\
 MAGERR\_R  & mag  &  $r$-band magnitude error \\
 MAGERR\_I  & mag  &  $i$-band magnitude error \\
 MAGERR\_Z  & mag  & $z$-band magnitude error \\
 MAGERR\_Y  & mag  & $Y$-band magnitude error \\
 A\_IMAGE & pixel & Major axis size based on an isophotal model \\ 
 B\_IMAGE & pixel & Minor axis size based on an isophotal model \\
 THETA\_J2000 & degree & Position angle of source in J2000 coordinates \\
 ERRA\_IMAGE & pixel & Error of major axis size based on an isophotal model \\
 ERRB\_IMAGE & pixel & Error of minor axis size based on an isophotal model \\
 ERRTHETA\_IMAGE & degree & Error of position angle of source  \\
 KRON\_RADIUS & pixel & Kron radius measured from detection image \\
 FLUX\_RADIUS\_I & pixel & Half-light radius for the object in $i$ band \\
 N\_NEIGHBOUR & & Number of neighbours used for photo-z estimation \\
 N\_FILTER   & &  Number of filters used \\
 PHOTO\_Z   & &  Estimated photometric redshift \\
 PHOTO\_ZERR & & Estimated photometric redshift error \\
 SPEC\_Z  & &    Spectroscopic redshift if available \\
 MEAN\_Z   & &   Mean spectroscopic redshift of $k$ nearest neighbours \\
 SIGMA\_Z   & &   Standard deviation of the spectroscopic redshifts of $k$ nearest neighbours \\
 NEAREST\_Z  & &  Spectroscopic redshift of the nearest neighbour \\
 MEAN\_DIS  & &  Mean Euclidean distance in the color space of neighbours \\
 CHI\_BEST  & &   $\chi^2$ of the best BC03 model fitting \\
 MOD\_BEST & & ID of the best-fitted BC03 model \\
 EBV\_BEST & mag &  Best estimated E(B-V) in three values of 0, 0.5, and 1.0 mag \\
 SCALE\_BEST & & Scaling factor between observed SED and model SED \\
 MAG\_ABS\_G & mag & $g$-band absolute magnitude \\
 MAG\_ABS\_R & mag & $r$-band absolute magnitude \\
 MAG\_ABS\_I & mag & $i$-band absolute magnitude \\
 MAG\_ABS\_Z & mag & $z$-band absolute magnitude \\
 MAG\_ABS\_Y & mag & $Y$-band absolute magnitude \\
 AGE\_BEST & yr &  Best estimated age \\
 AGE\_INF & yr & Lower limit of age with 68\% confidence level \\
 AGE\_MED  & yr &   Median age \\
 AGE\_SUP & yr &    Upper limit of age with 68\% confidence level \\
 MASS\_BEST & dex &  Best estimated log stellar mass ($M_{\sun}$) \\
 MASS\_INF  & dex &  Lower limit of log stellar mass with 68\% confidence level \\
 MASS\_MED & dex &   Median log stellar mass \\
 MASS\_SUP  & dex &  Upper limit of log stellar mass with 68\% confidence level \\
 SFR\_BEST & $M_{\sun}$ yr$^{-1}$ &   Best estimated SFR \\
 SFR\_INF   & $M_{\sun}$ yr$^{-1}$ &  Lower limit of SFR with 68\% confidence level \\
 SFR\_MED & $M_{\sun}$ yr$^{-1}$ &  Median SFR \\
 SFR\_SUP  & $M_{\sun}$ yr$^{-1}$ &  Upper limit of SFR with 68\% confidence level \\
 SSFR\_BEST & yr$^{-1}$ &  Best estimated SSFR \\
 SSFR\_INF & yr$^{-1}$ & Lower limit of SSFR with 68\%confidence level \\
 SSFR\_MED  & yr$^{-1}$ &   Median SSFR \\
 SSFR\_SUP  & yr$^{-1}$ &   Upper limit of SSFR with 68\%confidence level \\
 LUM\_NUV\_BEST & dex &  Best estimated log NUV luminosity ($L_{\sun}$) \\
 LUM\_R\_BEST & dex &   Best estimated log $R$-band luminosity ($L_{\sun}$) \\
 LUM\_K\_BEST & dex &  Best estimated log $K$-band luminosity ($L_{\sun}$) \\
\hline
\end{longtable}
\end{center}

\begin{center}
\begin{longtable}{lll}
\caption{Column description of the photo-z catalog for HSC-SSP} \label{tab:hsc_pz}\\
 \hline
  \hline
 Column & Unit  & Description \\
  \hline
  \endfirsthead
 \multicolumn{3}{c}%
{\tablename\ \thetable\ -- \textit{Continued from previous page}} \\
\hline
 \hline
Column & Unit  & Description \\
\hline
\endhead
\hline \multicolumn{3}{r}{\textit{Continued on next page}} \\
\endfoot
\hline
\endlastfoot
 OBJECT\_ID &  & Object ID in HSC-SSP \\
 RA  & degree &  R.A. in J2000  \\
 DEC  & degree & Declination in J2000 \\
 MAG\_G  & mag  &  $g$-band magnitude \\
 MAG\_R  & mag   &  $r$-band magnitude \\
 MAG\_I  & mag    &  $u$-band magnitude \\
 MAG\_Z  & mag &   $z$-band magnitude \\
 MAG\_Y  & mag  &  $y$-band magnitude \\
 MAGERR\_G  & mag  &  $g$-band magnitude error \\
 MAGERR\_R  & mag  &  $r$-band magnitude error \\
 MAGERR\_I  & mag  &  $i$-band magnitude error \\
 MAGERR\_Z  & mag  & $z$-band magnitude error \\
 MAGERR\_Y  & mag  & $y$-band magnitude error \\
 E11\_I & arcsec$^2$ & weighted average of ellipse component in $i$ band \\
 E12\_I & arcsec$^2$ & weighted average of ellipse component in $i$ band \\
 E22\_I & arcsec$^2$ & weighted average of ellipse component in $i$ band \\
 FRACDEV\_I &  & fraction of flux for de Vaucouleur component in $i$ band \\
 N\_NEIGHBOUR & & Number of neighbours used for photo-z estimation \\
 N\_FILTER   & &  Number of filters used \\
 PHOTO\_Z   & &  Estimated photometric redshift \\
 PHOTO\_ZERR & & Estimated photometric redshift error \\
 SPEC\_Z  & &    Spectroscopic redshift if available \\
 MEAN\_Z   & &   Mean spectroscopic redshift of $k$ nearest neighbours \\
 SIGMA\_Z   & &   Standard deviation of the spectroscopic redshifts of $k$ nearest neighbours \\
 NEAREST\_Z  & &  Spectroscopic redshift of the nearest neighbour \\
 MEAN\_DIS  & &  Mean Euclidean distance in the color space of neighbours \\
 CHI\_BEST  & &   $\chi^2$ of the best BC03 model fitting \\
 MOD\_BEST & & ID of the best-fitted BC03 model \\
 EBV\_BEST & mag &  Best estimated E(B-V) in three values of 0, 0.5, and 1.0 mag \\
 SCALE\_BEST & & Scaling factor between observed SED and model SED \\
 MAG\_ABS\_G & mag & $g$-band absolute magnitude \\
 MAG\_ABS\_R & mag & $r$-band absolute magnitude \\
 MAG\_ABS\_I & mag & $i$-band absolute magnitude \\
 MAG\_ABS\_Z & mag & $z$-band absolute magnitude \\
 MAG\_ABS\_Y & mag & $y$-band absolute magnitude \\
 AGE\_BEST & yr &   Best estimated age \\
 AGE\_INF & yr &    Lower limit of age with 68\% confidence level \\
 AGE\_MED  & yr &   Median age \\
 AGE\_SUP & yr &    Upper limit of age with 68\% confidence level \\
 MASS\_BEST & dex &  Best estimated log stellar mass ($M_{\sun}$) \\
 MASS\_INF  & dex &  Lower limit of log stellar mass with 68\% confidence level \\
 MASS\_MED & dex &   Median log stellar mass \\
 MASS\_SUP  & dex &  Upper limit of log stellar mass with 68\% confidence level \\
 SFR\_BEST & $M_{\sun}$ yr$^{-1}$ &   Best estimated SFR \\
 SFR\_INF   & $M_{\sun}$ yr$^{-1}$ &  Lower limit of SFR with 68\% confidence level \\
 SFR\_MED & $M_{\sun}$ yr$^{-1}$ &  Median SFR \\
 SFR\_SUP  & $M_{\sun}$ yr$^{-1}$ &  Upper limit of SFR with 68\% confidence level \\
 SSFR\_BEST & yr$^{-1}$ &  Best estimated SSFR \\
 SSFR\_INF & yr$^{-1}$ & Lower limit of SSFR with 68\%confidence level \\
 SSFR\_MED  & yr$^{-1}$ &   Median SSFR \\
 SSFR\_SUP  & yr$^{-1}$ &   Upper limit of SSFR with 68\%confidence level \\
 LUM\_NUV\_BEST & dex &  Best estimated log NUV luminosity ($L_{\sun}$) \\
 LUM\_R\_BEST & dex &   Best estimated log R-band luminosity ($L_{\sun}$) \\
 LUM\_K\_BEST & dex &  Best estimated log K-band luminosity ($L_{\sun}$) \\
\hline
\end{longtable}
\end{center}
\normalsize

\section{The cluster catalogs}
Table \ref{tab:desi_cluster}, \ref{tab:des_cluster}, and \ref{tab:hsc_cluster} list the the content in our cluster catalogs for DESI, DES, and HSC-SSP, respectively. 

\begin{center}
\begin{longtable}{lll}
\caption{Column description of the cluster catalog for DESI} \label{tab:desi_cluster} \\
 \hline
  \hline
 Column & Unit  & Description \\
  \hline
  \endfirsthead
 \multicolumn{3}{c}%
{\tablename\ \thetable\ -- \textit{Continued from previous page}} \\
\hline
 \hline
Column & Unit  & Description \\
\hline
\endhead
\hline \multicolumn{3}{r}{\textit{Continued on next page}} \\
\endfoot
\hline
\endlastfoot
 CLUSTER\_ID &  & Cluster ID \\
 RA\_PEAK & degree &  R.A. for the density peak (J2000) \\
 DEC\_PEAK & degree & decl. for the density peak (J2000) \\
 PHOTO\_Z\_PEAK &  & Photometric redshift for the density peak \\
 SPEC\_Z\_PEAK &   & Spectroscopic redshift for the density peak if existing \\
 LOC\_DEN\_PEAK &  & Local density for the density  peak \\
 LOC\_BKG\_PEAK &   & Local background density for the density peak \\
 N\_1MPC &  & Number of member galaxies within 1 Mpc from the cluster center \\
 L\_1MPC & $L^*$  & Total luminosity of member galaxies within 1 Mpc from the cluster center \\
 M\_500 & $\log_{10}(M_\odot)$ & Total mass of the cluster $M_{500}$ \\
 R\_500 & Mpc & Characteristic radius $R_{500}$ \\
 RICHNESS &  &  Cluster richness that is equal to L\_1MPC \\
 ID\_BCG &  & Object ID for the BCG \\ 
 RA\_BCG & degree  & R.A. for the BCG (J2000) for the BCG \\
 DEC\_BCG & degree  & decl. for the BCG (J2000) for the BCG \\
 PHOTO\_Z\_BCG &  & Photometric redshift for the BCG \\
 PHOTO\_ZERR\_BCG &  & Photometric redshift error for the BCG \\ 
 SPEC\_Z\_BCG &  & Spectroscopic redshift for the BCG if existing \\ 
 MAG\_G\_BCG & mag  & $g$-band magnitude for the BCG \\
 MAG\_R\_BCG &  mag & $r$-band magnitude for the BCG \\
 MAG\_Z\_BCG &  mag & $z$-band magnitude for the BCG \\
 MAG\_W1\_BCG &  mag  & $W1$-band magnitude for the BCG \\
 MAG\_W2\_BCG &  mag  & $W2$-band magnitude for the BCG \\
 MAGERR\_G\_BCG &  mag  & $g$-band magnitude error for the BCG \\
 MAGERR\_R\_BCG &  mag  & $r$-band magnitude error for the BCG \\ 
 MAGERR\_Z\_BCG &  mag  & $z$-band magnitude error for the BCG \\ 
 MAGERR\_W1\_BCG &  mag  & $W1$-band magnitude error for the BCG \\
 MAGERR\_W2\_BCG &  mag  & $W2$-band magnitude error for the BCG \\ 
 GALDEPTH\_G\_BCG  & mag &  $5\sigma$ galaxy depth in $g$ band for the BCG \\
 GALDEPTH\_R\_BCG & mag &  $5\sigma$ galaxy depth in $r$ band for the BCG \\
 GALDEPTH\_Z\_BCG  & mag &  $5\sigma$ galaxy depth in $z$ band for the BCG \\
 TYPE\_BCG &  & Morphological type for the BCG \\
 SHAPE\_R\_BCG & arcsec & Half-light radius of galaxy model for the BCG \\
 SHAPE\_R\_IVAR\_BCG & 1/arcsec$^2$ & Inverse variance of SHAPE\_R for the BCG \\
 SHAPE\_E1\_BCG &  & Ellipticity component 1 of galaxy model for the BCG \\
 SHAPE\_E1\_IVAR\_BCG &  & Inverse variance of SHAPE\_E1 for the BCG \\
 SHAPE\_E2\_BCG &  & Ellipticity component 2 of galaxy model for the BCG \\
 SHAPE\_E2\_IVAR\_BCG &  & Inverse variance of SHAPE\_E2 for the BCG \\
 SERSIC\_BCG &  & Power-law index of the Sersic profile model for the BCG \\
 SERSIC\_IVAR\_BCG &  & Inverse variance of SERSIC for the BCG \\
 MAG\_ABS\_G\_BCG &  mag & $g$-band absolute magnitude for the BCG \\ 
 MAG\_ABS\_R\_BCG &  mag  & $r$-band absolute magnitude for the BCG \\ 
 MAG\_ABS\_Z\_BCG &  mag  & $z$-band absolute magnitude for the BCG \\ 
 MAG\_ABS\_W1\_BCG &  mag  & $W1$-band absolute magnitude for the BCG \\
 MAG\_ABS\_W2\_BCG &  mag  & $W2$-band absolute magnitude for the BCG \\ 
 AGE\_BEST\_BCG & yr &   Best estimated age for the BCG \\
 AGE\_INF\_BCG & yr &    Lower limit of age with 68\% confidence level for the BCG \\
 AGE\_MED\_BCG  & yr &   Median age for the BCG \\
 AGE\_SUP\_BCG & yr &    Upper limit of age with 68\% confidence level for the BCG \\
 MASS\_BEST\_BCG & dex &  Best estimated log stellar mass ($M_{\sun}$) for the BCG \\
 MASS\_INF\_BCG  & dex &  Lower limit of log stellar mass with 68\% confidence level for the BCG \\
 MASS\_MED\_BCG & dex &   Median log stellar mass for the BCG \\
 MASS\_SUP\_BCG  & dex &  Upper limit of log stellar mass with 68\% confidence level for the BCG \\
 SFR\_BEST\_BCG & $M_{\sun}$ yr$^{-1}$ &   Best estimated SFR for the BCG \\
 SFR\_INF\_BCG   & $M_{\sun}$ yr$^{-1}$ &  Lower limit of SFR with 68\% confidence level for the BCG \\
 SFR\_MED\_BCG & $M_{\sun}$ yr$^{-1}$ &  Median SFR for the BCG \\
 SFR\_SUP\_BCG  & $M_{\sun}$ yr$^{-1}$ &  Upper limit of SFR with 68\% confidence level for the BCG \\
 SSFR\_BEST\_BCG & yr$^{-1}$ &  Best estimated SSFR for the BCG \\
 SSFR\_INF\_BCG & yr$^{-1}$ & Lower limit of SSFR with 68\%confidence level for the BCG \\
 SSFR\_MED\_BCG  & yr$^{-1}$ &   Median SSFR for the BCG \\
 SSFR\_SUP\_BCG  & yr$^{-1}$ &   Upper limit of SSFR with 68\%confidence level for the BCG \\
 LUM\_NUV\_BEST\_BCG & dex &  Best estimated log NUV luminosity ($L_{\sun}$) for the BCG \\
 LUM\_R\_BEST\_BCG & dex &   Best estimated log R-band luminosity ($L_{\sun}$) for the BCG \\
 LUM\_K\_BEST\_BCG & dex &  Best estimated log K-band luminosity ($L_{\sun}$) for the BCG \\ 
 \hline
\end{longtable}
\end{center}

\footnotesize
\begin{center}
\begin{longtable}{lll}
\caption{Column description of the cluster catalog for DES} \label{tab:des_cluster} \\
 \hline
  \hline
 Column & Unit  & Description \\
  \hline
  \endfirsthead
 \multicolumn{3}{c}%
{\tablename\ \thetable\ -- \textit{Continued from previous page}} \\
\hline
 \hline
Column & Unit  & Description \\
\hline
\endhead
\hline \multicolumn{3}{r}{\textit{Continued on next page}} \\
\endfoot
\hline
\endlastfoot
 CLUSTER\_ID &  & Cluster ID \\
 RA\_PEAK & degree &  R.A. for the density peak (J2000) \\
 DEC\_PEAK & degree & decl. for the density peak (J2000) \\
 PHOTO\_Z\_PEAK &  & Photometric redshift for the density peak \\
 SPEC\_Z\_PEAK &  & Spectroscopic redshift for the density peak if existing \\
 LOC\_DEN\_PEAK &  & Local density for the density  peak \\
 LOC\_BKG\_PEAK &   & Local background density for the density peak \\
 N\_1MPC &  & Number of member galaxies within 1 Mpc from the cluster center \\
 L\_1MPC & $L^*$  & Total luminosity of member galaxies within 1 Mpc from the cluster center \\
 M\_500 & $\log_{10}(M_\odot)$ & Total mass of the cluster $M_{500}$ \\
 R\_500 & Mpc & Characteristic radius $R_{500}$ \\
 RICHNESS &  &  Cluster richness that is equal to L\_1MPC \\
 ID\_BCG &  & Object ID for the BCG \\ 
 RA\_BCG & degree & R.A. for the BCG (J2000) \\
 DEC\_BCG & degree & decl. for the BCG (J2000) \\
 PHOTO\_Z\_BCG & & Photometric redshift for the BCG \\
 PHOTO\_ZERR\_BCG &  & Photometric redshift error for the BCG \\ 
 SPEC\_Z\_BCG &  & Spectroscopic redshift for the BCG if existing \\ 
 MAG\_G\_BCG & mag & $g$-band magnitude for the BCG \\
 MAG\_R\_BCG &  mag  & $r$-band magnitude for the BCG \\
 MAG\_I\_BCG &  mag & $i$-band magnitude for the BCG \\
 MAG\_Z\_BCG &  mag  & $z$-band magnitude for the BCG \\
 MAG\_Y\_BCG &  mag & $Y$-band magnitude for the BCG \\
 MAGERR\_G\_BCG &  mag & $g$-band magnitude error for the BCG \\
 MAGERR\_R\_BCG &  mag & $r$-band magnitude error for the BCG \\ 
 MAGERR\_I\_BCG &  mag & $i$-band magnitude error for the BCG \\ 
 MAGERR\_Z\_BCG &  mag & $z$-band magnitude error for the BCG \\
 MAGERR\_Y\_BCG &  mag & $Y$-band magnitude error for the BCG \\ 
 A\_IMAGE\_BCG & pixel & Major axis size based on an isophotal model for the BCG \\ 
 B\_IMAGE\_BCG & pixel & Minor axis size based on an isophotal model for the BCG \\
 THETA\_J2000\_BCG & degree & Position angle of source in J2000 coordinates for the BCG \\
 ERRA\_IMAGE\_BCG & pixel & Error of major axis size based on an isophotal model for the BCG \\
 ERRB\_IMAGE\_BCG & pixel & Error of minor axis size based on an isophotal model for the BCG \\
 ERRTHETA\_IMAGE\_BCG & degree & Error of position angle of source for the BCG  \\
 KRON\_RADIUS\_BCG & pixel & Kron radius measured from detection image for the BCG \\
 FLUX\_RADIUS\_I\_BCG & pixel & Half-light radius for the object in $i$ band for the BCG \\
 MAG\_ABS\_G\_BCG &  mag & $g$-band absolute magnitude for the BCG \\ 
 MAG\_ABS\_R\_BCG &  mag & $r$-band absolute magnitude for the BCG \\ 
 MAG\_ABS\_I\_BCG &  mag & $i$-band absolute magnitude for the BCG \\ 
 MAG\_ABS\_Z\_BCG &  mag & $z$-band absolute magnitude for the BCG \\
 MAG\_ABS\_Y\_BCG &  mag & $Y$-band absolute magnitude for the BCG \\ 
 AGE\_BEST\_BCG & yr &   Best estimated age for the BCG \\
 AGE\_INF\_BCG & yr &    Lower limit of age with 68\% confidence level for the BCG \\
 AGE\_MED\_BCG  & yr &   Median age for the BCG \\
 AGE\_SUP\_BCG & yr &    Upper limit of age with 68\% confidence level for the BCG \\
 MASS\_BEST\_BCG & dex &  Best estimated log stellar mass ($M_{\sun}$) for the BCG \\
 MASS\_INF\_BCG  & dex &  Lower limit of log stellar mass with 68\% confidence level for the BCG \\
 MASS\_MED\_BCG & dex &   Median log stellar mass for the BCG \\
 MASS\_SUP\_BCG  & dex &  Upper limit of log stellar mass with 68\% confidence level for the BCG \\
 SFR\_BEST\_BCG & $M_{\sun}$ yr$^{-1}$ &   Best estimated SFR for the BCG \\
 SFR\_INF\_BCG   & $M_{\sun}$ yr$^{-1}$ &  Lower limit of SFR with 68\% confidence level for the BCG \\
 SFR\_MED\_BCG & $M_{\sun}$ yr$^{-1}$ &  Median SFR for the BCG \\
 SFR\_SUP\_BCG  & $M_{\sun}$ yr$^{-1}$ &  Upper limit of SFR with 68\% confidence level for the BCG \\
 SSFR\_BEST\_BCG & yr$^{-1}$ &  Best estimated SSFR for the BCG \\
 SSFR\_INF\_BCG & yr$^{-1}$ & Lower limit of SSFR with 68\%confidence level for the BCG \\
 SSFR\_MED\_BCG  & yr$^{-1}$ &   Median SSFR for the BCG \\
 SSFR\_SUP\_BCG  & yr$^{-1}$ &   Upper limit of SSFR with 68\%confidence level for the BCG \\
 LUM\_NUV\_BEST\_BCG & dex &  Best estimated log NUV luminosity ($L_{\sun}$) for the BCG \\
 LUM\_R\_BEST\_BCG & dex &   Best estimated log R-band luminosity ($L_{\sun}$) for the BCG \\
 LUM\_K\_BEST\_BCG & dex &  Best estimated log K-band luminosity ($L_{\sun}$) for the BCG \\
 \hline
\end{longtable}
\end{center}

 \begin{center}
\begin{longtable}{lll}
\caption{Column description of the cluster catalog for HSC-SSP} \label{tab:hsc_cluster} \\
 \hline
  \hline
 Column & Unit  & Description \\
  \hline
  \endfirsthead
 \multicolumn{3}{c}%
{\tablename\ \thetable\ -- \textit{Continued from previous page}} \\
\hline
 \hline
Column & Unit  & Description \\
\hline
\endhead
\hline \multicolumn{3}{r}{\textit{Continued on next page}} \\
\endfoot
\hline
\endlastfoot
 CLUSTER\_ID &  & Cluster ID \\
 RA\_PEAK & degree &  R.A. for the density peak (J2000) \\
 DEC\_PEAK & degree & decl. for the density peak (J2000) \\
 PHOTO\_Z\_PEAK &  & Photometric redshift for the density peak \\
 SPEC\_Z\_PEAK &  & Spectroscopic redshift for the density peak if existing \\
 LOC\_DEN\_PEAK &  & Local density for the density  peak \\
 LOC\_BKG\_PEAK &   & Local background density for the density peak \\
 N\_1MPC &  & Number of member galaxies within 1 Mpc from the cluster center \\
 L\_1MPC & $L^*$  & Total luminosity of member galaxies within 1 Mpc from the cluster center \\
 M\_500 & $\log_{10}(M_\odot)$ & Total mass of the cluster $M_{500}$ \\
 R\_500 & Mpc &  Characteristic radius $R_{500}$ \\
 RICHNESS &  & Cluster richness that is equal to L\_1MPC \\
 ID\_BCG &  & Object ID for the BCG \\ 
 RA\_BCG & degree & R.A. for the BCG (J2000) for the BCG \\
 DEC\_BCG & degree & decl. for the BCG (J2000) for the BCG \\
 PHOTO\_Z\_BCG &  & Photometric redshift for the BCG \\
 PHOTO\_ZERR\_BCG &  & Photometric redshift error for the BCG \\ 
 SPEC\_Z\_BCG &  & Spectroscopic redshift for the BCG if existing \\ 
 MAG\_G\_BCG & mag & $g$-band magnitude for the BCG \\
 MAG\_R\_BCG &  mag & $r$-band magnitude for the BCG \\
 MAG\_I\_BCG &  mag  & $i$-band magnitude for the BCG \\
 MAG\_Z\_BCG &  mag & $z$-band magnitude for the BCG \\
 MAG\_Y\_BCG &  mag & $y$-band magnitude for the BCG \\
 MAGERR\_G\_BCG &  mag  & $g$-band magnitude error for the BCG \\
 MAGERR\_R\_BCG &  mag  & $r$-band magnitude error for the BCG \\ 
 MAGERR\_I\_BCG &  mag  & $i$-band magnitude error for the BCG \\ 
 MAGERR\_Z\_BCG &  mag  & $z$-band magnitude error for the BCG \\
 MAGERR\_Y\_BCG &  mag  & $y$-band magnitude error for the BCG \\ 
 E11\_I\_BCG & arcsec$^2$ & weighted average of ellipse component in $i$ band for the BCG \\
 E12\_I\_BCG & arcsec$^2$ & weighted average of ellipse component in $i$ band for the BCG \\
 E22\_I\_BCG & arcsec$^2$ & weighted average of ellipse component in $i$ band for the BCG \\
 FRACDEV\_I\_BCG &  & fraction of flux for de Vaucouleur component in $i$ band for the BCG \\
 MAG\_ABS\_G\_BCG &  mag & $g$-band absolute magnitude for the BCG \\ 
 MAG\_ABS\_R\_BCG &  mag  & $r$-band absolute magnitude for the BCG \\ 
 MAG\_ABS\_I\_BCG &  mag & $i$-band absolute magnitude for the BCG \\ 
 MAG\_ABS\_Z\_BCG &  mag & $z$-band absolute magnitude for the BCG \\
 MAG\_ABS\_Y\_BCG &  mag  & $y$-band absolute magnitude for the BCG \\ 
 AGE\_BEST\_BCG & yr &   Best estimated age for the BCG \\
 AGE\_INF\_BCG & yr &    Lower limit of age with 68\% confidence level for the BCG \\
 AGE\_MED\_BCG  & yr &   Median age for the BCG \\
 AGE\_SUP\_BCG & yr &    Upper limit of age with 68\% confidence level for the BCG \\
 MASS\_BEST\_BCG & $\log_{10}(M_\odot)$ & Logarithmic stellar mass for the BCG \\ 
 MASS\_INF\_BCG & $\log_{10}(M_\odot)$  & Lower limit of logarithmic stellar mass with 68\% confidence level for the BCG \\
 MASS\_MED\_BCG & dex &   Median log stellar mass for the BCG \\
 MASS\_SUP\_BCG & $\log_{10}(M_\odot)$ & Upper limit of logarithmic stellar mass with 68\% confidence level for the BCG \\
 SFR\_BEST\_BCG & $M_{\sun}$ yr$^{-1}$ &   Best estimated SFR for the BCG \\
 SFR\_INF\_BCG   & $M_{\sun}$ yr$^{-1}$ &  Lower limit of SFR with 68\% confidence level for the BCG \\
 SFR\_MED\_BCG & $M_{\sun}$ yr$^{-1}$ &  Median SFR for the BCG \\
 SFR\_SUP\_BCG  & $M_{\sun}$ yr$^{-1}$ &  Upper limit of SFR with 68\% confidence level for the BCG \\
 SSFR\_BEST\_BCG & yr$^{-1}$ &  Best estimated SSFR for the BCG \\
 SSFR\_INF\_BCG & yr$^{-1}$ & Lower limit of SSFR with 68\%confidence level for the BCG \\
 SSFR\_MED\_BCG  & yr$^{-1}$ &   Median SSFR for the BCG \\
 SSFR\_SUP\_BCG  & yr$^{-1}$ &   Upper limit of SSFR with 68\%confidence level for the BCG \\
 LUM\_NUV\_BEST\_BCG & dex &  Best estimated log NUV luminosity ($L_{\sun}$) for the BCG \\
 LUM\_R\_BEST\_BCG & dex &   Best estimated log R-band luminosity ($L_{\sun}$) for the BCG \\
 LUM\_K\_BEST\_BCG & dex &  Best estimated log K-band luminosity ($L_{\sun}$) for the BCG \\
 \hline
\end{longtable}
\end{center}

\label{lastpage}

\end{document}